\documentstyle[]{IEEEtran}

\newcommand{\afh}{\mbox{$ Ar/CF_4/CH_4 $}}
\newcommand{\afo}{\mbox{$ Ar/CF_4/CO_2 $}}

\begin{document}

\begin{titlepage}

\pagebreak

\vspace*{0.1cm}
\begin{center}
{\huge Institute of Theoretical }
\end{center} 
\begin{center}
{\huge and Experimental Physics }
\end{center} 
\begin{center}
\vspace{0.5cm}
{\huge Preprint  43-00 }
\end{center} 
\vspace{3.0cm}
\begin{center}
{\large \bf M.Danilov, L.Laptin, I.Tichomirov, M.Titov, Yu.Zaitsev}
\end{center} 

\begin{center}
\vspace{2.cm}
{\large\bf Aging tests of the proportional wire chambers}
\end{center}
\begin{center}
{\large\bf using $\afh$(74:20:6), $\afh$(67:30:3)}
\end{center}
\begin{center}
{\large\bf and $\afo$(65:30:5) mixtures}
\end{center}
\begin{center}
{\large\bf for the HERA-B Muon Detector}
\end{center}
 
\vspace{7.5cm}

\begin{center}
{\large\bf Moscow 2000}
\end{center} 

\end{titlepage}

\newpage


\begin{abstract}
\large 

The Muon Detector of the HERA-B experiment at DESY is a gaseous 
detector that provides muon identification in a high-rate hadronic
environment. 
We present our studies on the properties  of several fast gases,
$\afh$ (74:20:6), $\afh$ (67:30:3) and $\afo$ (65:30:5), 
which have been found to fulfill muon detection requirements.

The severe radiation environment of the HERA-B experiment leads to the maximum
charge deposit on a wire, within the muon detector, of 200 mC/cm per year.
For operation in such an environment, the main criteria for the choice of gas
turned out to be stability against aging.
  An overview of aging results from
laboratory setups and experimental detectors for binary and ternary 
mixtures of $Ar$, $CH_4$, $CF_4$ and $CO_2$ is presented and
the relevance of the various aging results is discussed.
Since it is not clear how to extrapolate aging results from small
to large areas of irradiation, the lifetime of 
aluminum proportional chambers was studied under various conditions.
In this paper we provide evidence
that aging results depend not only upon the 
total collected charge.   
It was found that the aging rate for irradiation with $Fe^{55}$
$X$-rays and 100~MeV $\alpha$-particles may differ by more than
two orders of magnitude.

\end{abstract}
\large
\clearpage

\pagenumbering{arabic}
\setcounter{page}{1}

\section {Introduction.}

 The HERA-B experiment is a hadronic B-factory at DESY, 
operating presently under LHC-like conditions, 
five years before startup of LHC.
The muon detection system in HERA-B is made up of three different types 
of gas proportional chambers: tube, pad and pixel~\cite{Proposal,TDR}. 
The readout of the chambers is based on the ASD-8
amplifier shaper discriminator chip developed by the University
of Pennsylvania~\cite{electronics}.
Strong restrictions on the gas choice are imposed by
the harsh operating conditions.
 In addition to the high radiation load, resulting in an  
accumulated charge on a wire of up to 200 mC/cm per year, the  
muon chambers have to withstand the presence of heavily 
ionizing particles, neutrons, and low energy gammas.
 The first level trigger  requires high hit efficiency and 
fast signal collection within the 96~ns time interval
between two consecutive bunch crossings.

In section 2, the requirements and limitations of the choice 
of gases for the muon system are explained and the chamber
characteristics studied with several fast $Ar/CF_4$-based mixtures
are discussed. 
 The overview of aging results from
laboratory setups and experimental detectors for binary 
and ternary mixtures of $Ar$, $CH_4$, $CO_2$ and $CF_4$ is 
presented in section 3.  
Since it is not clear how to extrapolate aging results 
from laboratory setups, when irradiating a small region of the wire,
to large areas of irradiation with hadronic particles, 
we have studied aging properties
of aluminum proportional chambers operated with $\afh$ and $\afo$
mixtures with $Fe^{55}$ and $Ru^{106}$ sources and in 
100~MeV $\alpha$-beam. 
The results of these aging tests performed in a variety of conditions
are summarized in sections 4 and 5.
 

\section{ Selection of fast gas mixtures.}

Large area gaseous detectors operated under 
high intensity particle fluxes impose 
new requirements for gas selection. 
The HERA-B muon system, with a total gas volume of 8~$m^3$,
puts rather stringent constraints
on the gaseous medium to be used: extremely low aging, 
high sparkproofness, good electrical properties (high drift 
velocity $v_d$, convenient operating electric field E/p),
good chemical properties (non-flammable, non-poisonous).
 Since it is necessary to replenish part of the gas to avoid 
contamination by air, the cost of the component gases is also an
important issue for gas selection.

For a minimum working voltage, noble gases are usually chosen
since they require the lowest electric field intensities
for avalanche formation.
 However, photons limit the attainable gas amplification  
in the $Ar$-operated counters to $\sim 10^{4}$
without encountering a permanent discharge induced
by the photon-feedback mechanism  taking place at the cathode.
 This problem can be remedied by the addition of gases,
such as hydrocarbons, alcohols, $CO_2$, $H_2 O$, $N H_3$.
  These molecular additives act like quenchers and improve 
the operational stability 
of the counter by absorption of ultraviolet photons from the 
multiplication process, de-excitation of $Ar$ atoms~\cite{scint},
and by the charge transfer 
mechanism prevent the neutralization of argon ions at the cathode,
thus reducing electron emission 
through predissociation of quench gas ions~\cite{start,co2}.
 Some of the polyatomic gases will also effectively increase 
the total ionization by the Penning effect.     

 Moreover, in pure $Ar$ the electron drift velocity is $\sim 30$
times lower than in $CH_4$, although their elastic cross 
sections are nearly identical. However, the addition of even
small quantities ($\sim$0.1$\%$) of a molecular gas increases the  
drift velocity in argon threefold and also alters the shape dependence
of $v_d$ on $E/P$~\cite{N2,book1}.
  In general, the drift velocity $v_d$ varies inversely with the 
product of the total scattering cross-section $\sigma_{sc}$
and the square root of the mean electron energy 
$\overline{\epsilon} $~\cite{fast1,fast2}.
 Traditionally, the fast gas mixtures were achieved by 
adding to $Ar$ a polyatomic gas~(usually a hydrocarbon or $CO_2$) 
which effectively shifts the low energy part of the 
electron distribution function into the region of the
Ramsauer-Townsend~(R-T) minimum, located at $\sim$ 0.5~eV,
of the momentum transfer and total electron scattering
cross-section of $Ar$.
The cooling of the electron swarm occurs as the energetic
electrons quickly transfer their energy to rotational and vibrational 
modes of the polyatomic gas, whose scattering cross-section has a
large inelastic component to the right of the R-T minimum. 
Furthermore, the cooling effect of a given gas in a high electric field 
gas amplification region need not be the same as it's cooling effect
in the lower electric field drift region. As was shown
in~\cite{armit}, the very similar effects observed 
in the $Ar/CH_4$ and $Ar/CO_2$ mixtures 
in the avalanche region are contrasted 
with different properties of these gases in the lower electric fields:
$CO_2$ is a much cooler gas than $CH_4$ in the drift region,
as is evidenced by the measured diffusion coefficients~\cite{cern}. 
These results open the
possibility of choosing a gas mixture to separately optimize
the desired drift and gain behavior in a particular application.

 Almost twenty years ago, $CF_4$ has been proposed for 
high rate environments as a most effective additive 
to raise the drift velocity in noble gases.
This gas obtains its large drift velocity because of the
sizeable Ramsauer-Townsend dip in the elastic cross-section which coincides
with a very large vibrational inelastic cross-section~\cite{biagi}.
 For the $Ar/CF_4$(90:10) mixture the 
electron drift velocity was found to exceed 10~cm/$\mu s$, 
which is twice as fast as for the conventional mixture
$Ar/CH_4$(90:10)~\cite{fast1}.
 Further increase of the $CF_4$ fraction 
in $Ar/CF_4$ mixtures tends to push 
the maximum drift velocity to the higher field region, 
but when the percentage of $CF_4$ becomes larger than 20~$\%$,
the $v_d$ is the same whatever the mixture,
indicating that it's mainly controlled by the $CF_4$ 
cross section~\cite{cf42,mix1,transport}.
 The lower vibrational modes, as well as  
the lower-lying negative ion states of $CF_4$ as compared to 
$CH_4$, may account for the superior drift properties 
of the $Ar/CF_4$ mixtures compared to  
$Ar/CH_4$~\cite{christ,book2}.

However, all binary $Ar/CF_4$ mixtures have rather poor energy 
resolution, compared to $Ar/CH_4$, due to the production 
of negative ions~\cite{fast1,attach}.
The dissociative attachment processes in $CF_4$ are the only mechanisms
leading to negative ion ($F^{-}$, $CF_3^{-}$) formation, via two
broad and overlapping resonances located between 4.5 and 
10~eV~\cite{cf42,cf41}, and can cause serious inefficiencies in the
collection of electrons and degrade the position resolution.
The main evidence for the electron attachment is the direct measurement
of the effective ionization coefficient $\overline{\alpha} = \alpha - \eta$;
where $\alpha$ and $\eta$ are, respectively, the electron impact ionization 
and electron attachment coefficients. The negative values 
of $\overline{\alpha}$ were observed in the drift region
in pure $CF_4$, $Ar/CF_4$(80:20) and
$Ar/CF_4/H_2 O$(80:18:2) mixtures. Very interestingly,
the electron thermalizing 
properties effected by substituting 10~$\%$ of $CF_4$ in the
$Ar/CF_4$(80:20)
by 10~$\%$ of $CO_2$ virtually eliminates the electron attachment
in $Ar/CF_4/CO_2$(80:10:10) mixture~\cite{christ}.

 Moreover, due to the small $CF_4$ quenching cross-sections 
of metastable $Ar$-states~\cite{quench1,quench2},
the $Ar/CF_4$ mixtures are not sufficiently
self-quenching and show an intolerable level of afterpulsing  
even at moderate gas amplifications~(see also section 2.1). 
 This is not the case for the $Ar$/hydrocarbon mixtures, where
no significant afterpulsing was observed so far at moderate gas gains.
At the same time, some of the $Ar/CO_2$ mixtures are known to suffer from a low 
sparkproofness~\cite{becker}, 
since the quenching ability of $CO_2$ is smaller than that of hydrocarbons.

\begin{figure}[ph]
\setlength{\unitlength}{1mm}
 \begin{picture}(160,200)
 \put(-10.0,80.0){\includegraphics{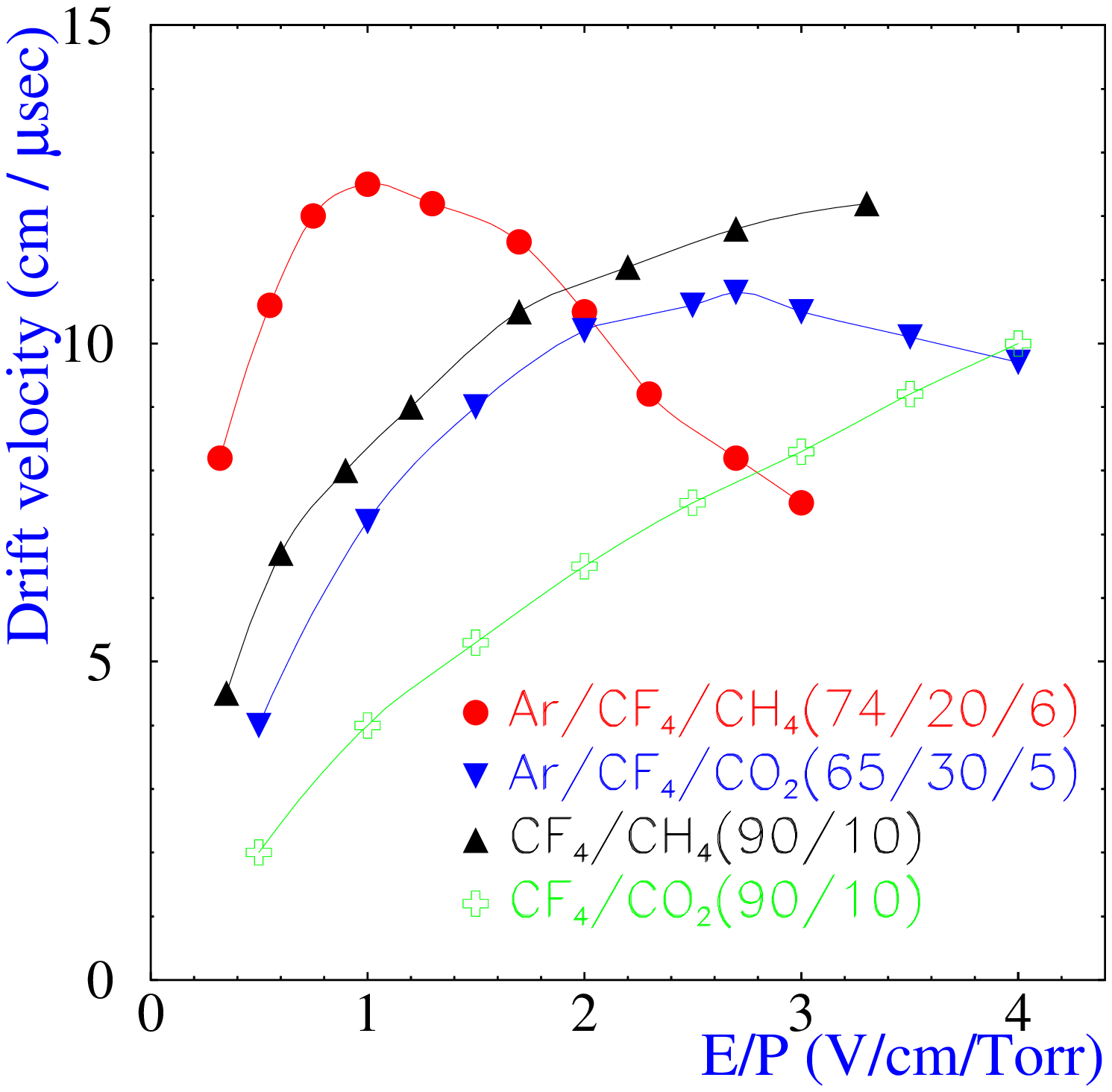} }
 \put(75.0,80.0){\includegraphics{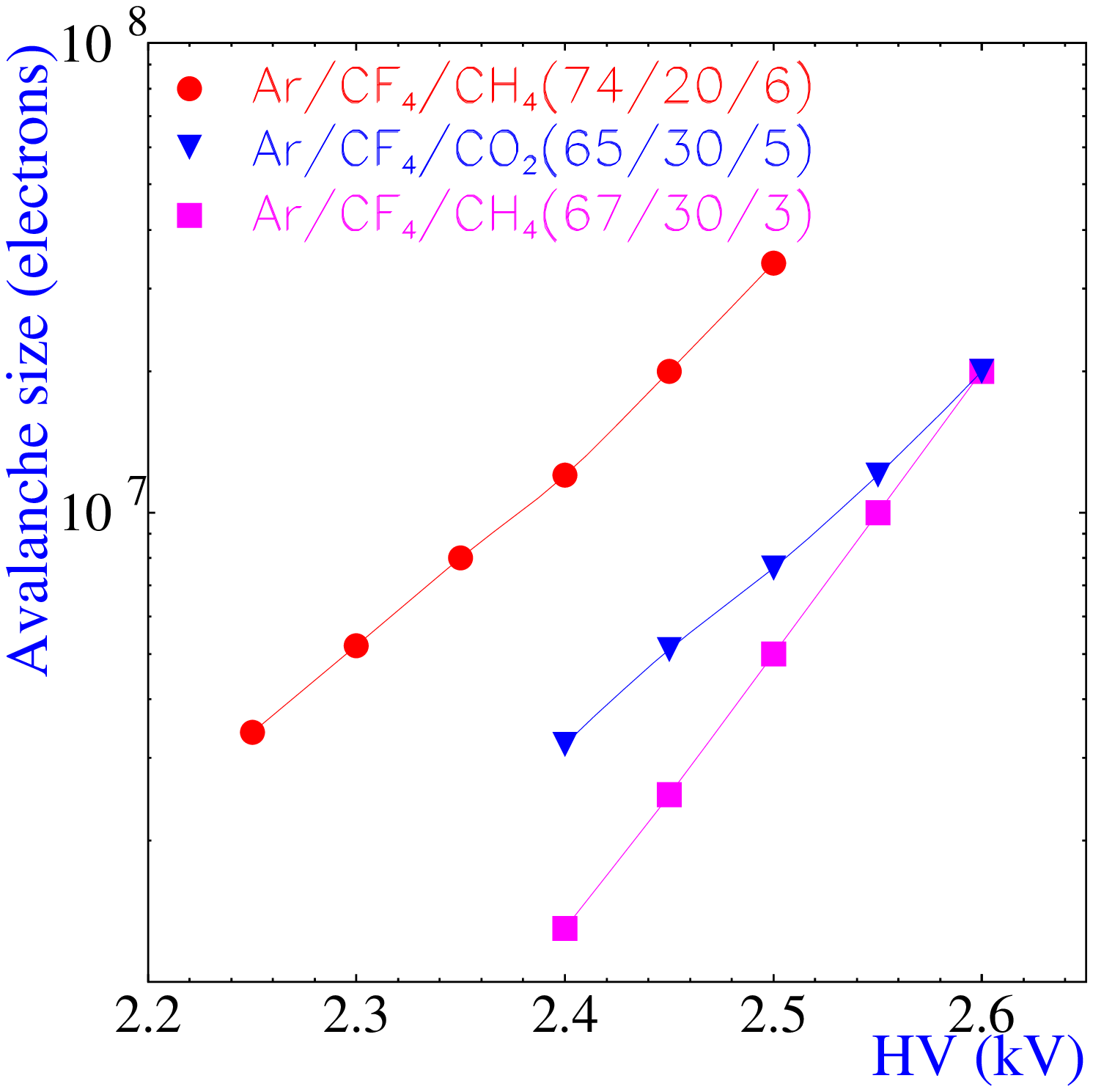} }
 \put(-10.0,-30.0){\includegraphics{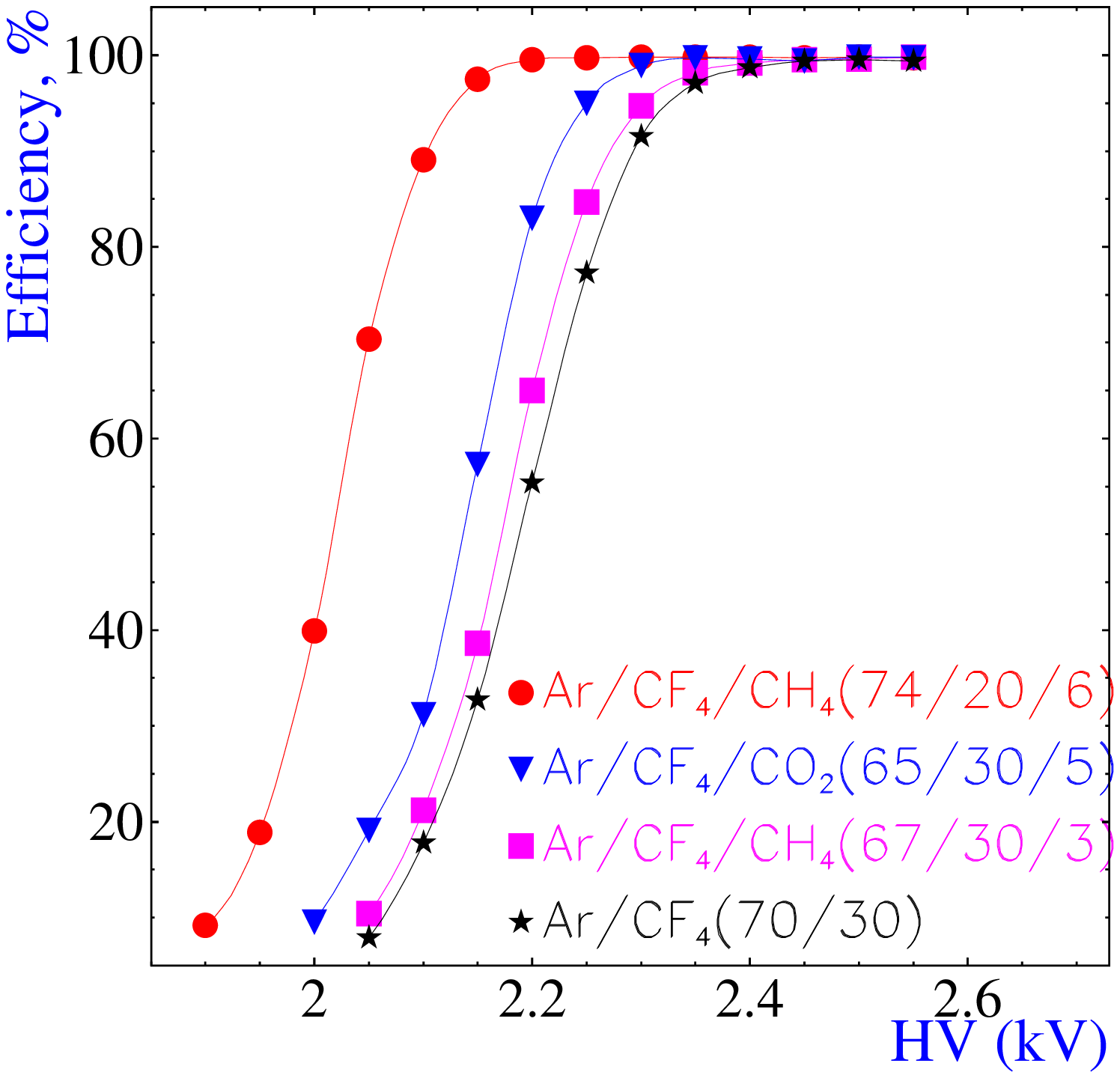} }
 \put(75.0,-30.0){\includegraphics{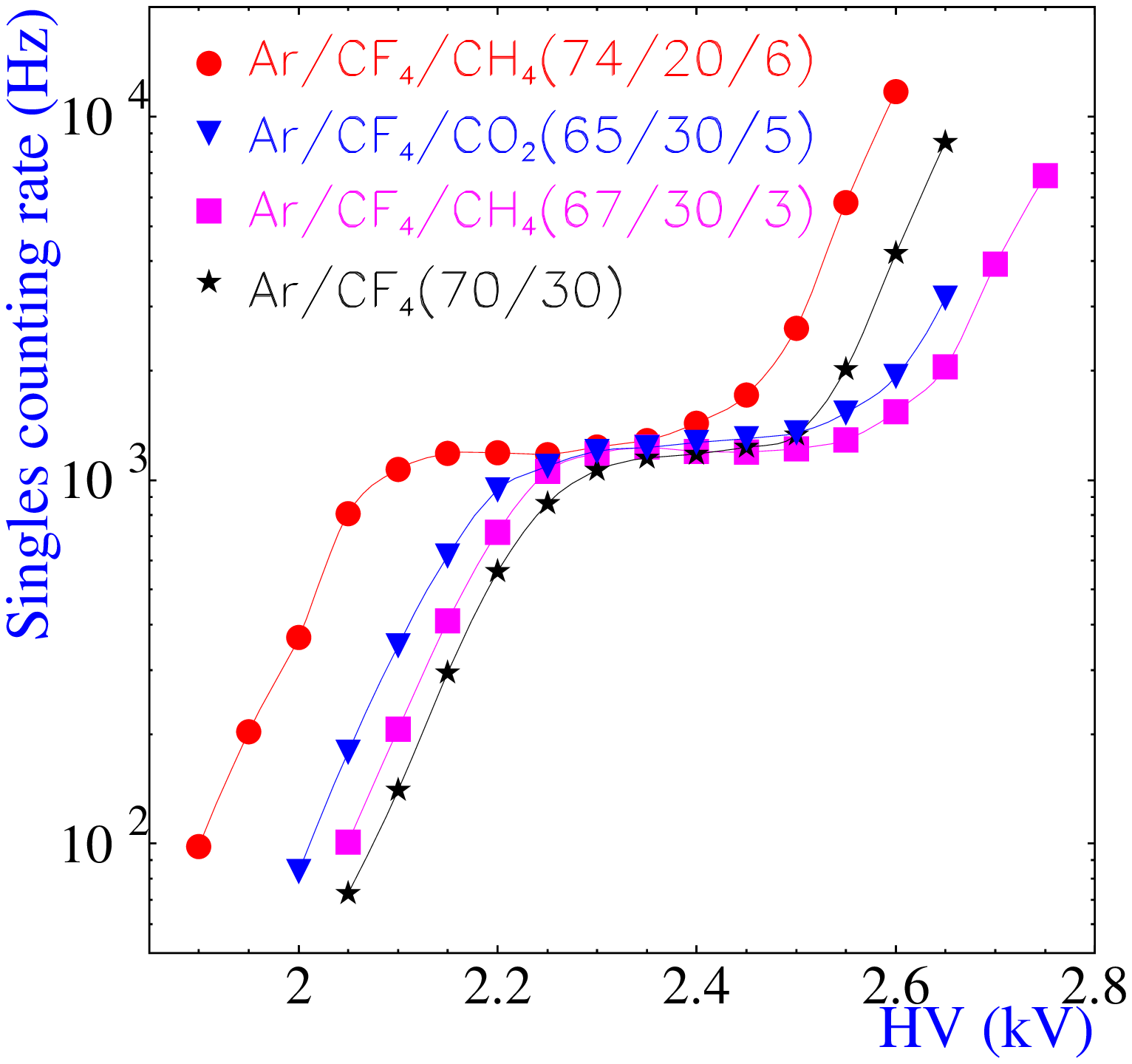} }
 \put(20,120){ Figure 1: Electron drift velocity }
 \put(30,114){ as a function of $E/P$ }
 \put(103,120){Figure 2: Avalanche size curves as a }
 \put(104,114){ function of HV for 5.9~keV $X$-rays }
 \put(20,10){Figure 3: Tube chamber efficiency }
 \put(35,4){ as a function of HV }
 \put(103,10){Figure 4: Singles counting rates for a}
 \put(103,4){ tube chambercell as a function of HV}
 \end{picture}
\end{figure}

The advantage of the enhanced drift velocity of $CF_4$ 
could be realized either in binary $CF_4$/quencher
or in ternary $Ar/CF_4$/quencher mixtures.
The choice of $CH_4$ among the hydrocarbons is 
dictated by the requirements that the drift velocity in the
lighter hydrocarbons is larger than that in the heavier  
ones~\cite{mix1}, especially in the  low-field region.
Fig. 1 shows the drift velocity for the $Ar/CF_4/CH_4$(74:20:6),
$CF_4/CH_4$(90:10), $Ar/CF_4/CO_2$(65:30:5),  $Ar/CF_4/CO_2$(80:10:10)
 and $CF_4/CO_2$(90:10) mixtures, as a function of the electric
field strength at atmospheric pressure. These results are taken from 
Refs.~\cite{christ,grimm,beck,mix2}. As can be seen from the curves,
the $CF_4/CH_4$ mixtures are significantly faster than 
$CF_4/CO_2$ for any given value of $E/P$.
  The use of $\afh$ could be more preferable for 
high-rate proportional chamber applications than $CF_4/CH_4$ mixtures, 
since the former shows fast rises in the drift velocity for 
the relatively low fields, with respect to the latter,
and the mobilities of $CH_4^+$ ions in $Ar$
are much higher than those in $CF_4$,
which is desirable in terms of space charge effects~\cite{mix1,mix2}.
 An additional disadvantage of the $CF_4/CH_4$ mixture is the necessity
to operate at high electric fields $E/P$ for a sufficient gas gain,
resulting in exceedingly high voltages in the chamber. 
This could introduce a problem  of electrostatic
instability of anode wires in large chambers.
Furthermore, considerations of cost and safety 
led us to consider $Ar/CF_4/CH_4$ mixtures as a
reasonable compromise between many experimental requirements.
 An optimization of the exact proportions of $Ar$, $CF_4$, and $CH_4$  in
the mixture was performed using the Magboltz and Garfield simulation programs.
As a result of simulation, $Ar/CF_4/CH_4$ (74:20:6) and another mixture
with a reduced $CH_4$ content $Ar/CF_4/CH_4$ (67:30:3) were chosen
for muon detector operation.
Unfortunately, mixtures with hydrocarbons have a tendency to 
cause polymerization effects, while
$CO_2$ quencher has been proven to give a much longer lifetimes.
However, addition of $CO_2$ appreciably changes the drift properties 
of the $Ar/CF_4/CO_2$ mixture, compared  
to the case of $Ar/CF_4/CH_4$;
reduces the maximum drift velocity, and shifts it to the higher 
field region.
In practice, we chose $Ar/CF_4/CO_2$(65:30:5) mixture, 
as a fall back gas, if the aging criterion becomes dominant.

Finally, in the  $Ar/CF_4/CH_4$ (74:20:6), 
$Ar/CF_4/CH_4$ (67:30:3) and $Ar/CF_4/CO_2$ (65:30:5) mixtures,
electron attachment processes may complicate gas performance.
However, the primary ionization in the chamber cell
is large enough and more than compensates for any losses  due to  
electron attachment, especially since only hit information is processed 
in the muon detector.


\subsection{ Tube Chamber Characteristics. }

 In this paper, we address the problem of gas selection 
and present the measured data only for tube 
chambers.
 The tube chamber is a closed-cell proportional wire chamber 
made from an aluminum profile~(wall thickness 2~mm)
with a drift cell 14~$\times$~12~mm$^2$ in cross section.
 Fig. 5 shows the schematic drawing of the tube chamber.
A gold-plated tungsten wire of 45~$\mu$m diameter and a length 
of nearly 3~m is stretched inside each cell and fixed 
mechanically with pins at the end of the chamber.
 In order to prevent efficiency losses due to dead spaces between cells, 
the tube chamber consists of two layers, each having 16 drift cells, 
shifted with respect to each other by half a cell size.

\begin{figure}[bth]
\setlength{\unitlength}{1mm}
 \begin{picture}(100,50)
\put(40.0,8.0){\includegraphics{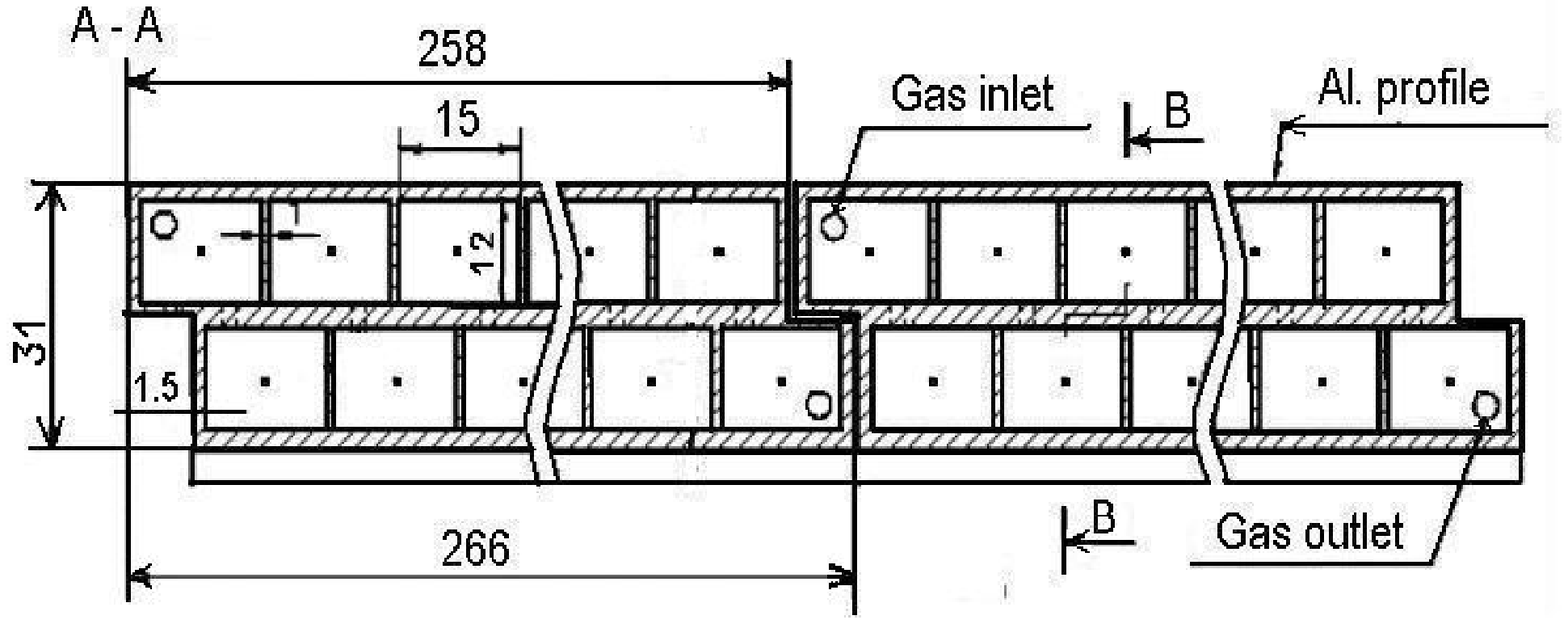}}
 \put(30,6){ Figure 5: Schematic drawing of the tube chamber. Dimensions are }
 \put(40,0){ indicated in units of mm. }
 \end{picture}
\end{figure}

 Firstly, we have collected the $Fe^{55}$ spectra for each mixture
under study: $\afh$ (74:20:6), $\afh$ (67:30:3) and $\afo$ (65:30:5)
with a goal of inferring the response to minimum ionizing particles.
Measured pulse heights were converted to avalanche sizes,
which correspond to the total number of electrons in the 
avalanche. The gas gain value
can be obtained  by dividing the observed pulse size by $\sim$ 200, 
the average number of primary electrons,
if the mixture has no electron attachment.
 Fig. 2 shows the total avalanche size increasing in an exponential 
manner with the high voltage, within the limits of the proportional region.
It has been shown previously~\cite{gain1} that the onset 
of secondary gas effects such as space charge saturation
may be detected as a break in the expected scaling 
of the gas gain with respect both to anode surface field and
to gradient of the electric field near the wire.
 It should be noted that for these data taken with an $Fe^{55}$
source with its fixed initial charge ($\sim$200 electrons),
a particular value of gas gain is equivalent to a definite value of the 
total charge in the avalanche. Space charge effects depend on the 
total avalanche size and distribution of the positive ion cloud. Hence,
these numerical results cannot be automatically applied to other 
ionization sources with different magnitudes or distributions of the
initial charge.

Secondly, the tube chamber characteristics have been 
studied on the 3~GeV electron test beam at DESY.
The beam intensity was $\sim$~1~$kHz/cm^{2}$.
 The trigger was initiated by a coincidence of 
four beam-defining scintillator counters,
between which the chamber was positioned.
Then, the number of events with at least one hit in the double layer
chamber, within the 96~ns, versus the number of triggered events
yields an efficiency.
Fig. 3 shows the measured efficiency 
for the mixtures being studied, as a function of high voltage. 
This figure demonstrates, that the actual efficiency is approximately 
100~$\%$ for the double layer chamber. 
However, for a single layer the usage of $\afo$(65:30:5) mixture  
results in about 7~$\%$ of signals registered outside the 96~ns  
(for perpendicular tracks), while in $\afh$ mixture nearly all
signals are collected within the 96~ns. 
 Fig. 4 presents singles (untriggered) counting rates for a 
tube chamber cell as a function of high voltage. 
From these curves, wide high voltage plateaus for a stable 
and efficient operation 
with  the $\afh$(74:20:6), $\afh$(67:30:3) and $\afo$(65:30:5) gases
are observed, where the counting rate is roughly uniform. 
Real afterpulses and multiple hits of ASD-8 per signal both 
contribute to the rise in singles rate curves, above the 
nominal beam intensity, with increasing  high voltage. 
 In comparison, $Ar/CF_4$(70:30) mixture
will not guarantee stable and efficient operation over
sufficient high-voltage range in the tube chambers.
The difference in singles rate curves, and correspondingly 
in detection efficiency,
between $Ar/CF_4$(70:30) and $\afh$(67:30:3) 
may be attributed to the enhanced gas electron production efficiency
and smaller electron attachment for the latter mixture.
In addition, onset of secondary avalanches in $Ar/CF_4$
limit the maximum operating voltage to 2.5~kV~(see Fig. 4), thus
causing a drastic decrease in the counting rate plateau width.
 
The results presented here justify 
the choice of $\afh$ (74:20:6), $\afh$ (67:30:3) and $\afo$ (65:30:5)
gases for muon detection with high efficiency.

\section{ General characteristics of aging processes. }

  Aging effects in proportional wire chambers,
a permanent degradation of operating
characteristics under sustained irradiation, has been and remains
the main limitation to their use in high-rate experiments~\cite{charpak}.
 Many processes are expected to occur simultaneously in the
gaseous discharges surrounding the wire
and it is nearly impossible to obtain
a quantitative chemical description in each particular case.
 There is much experimental information, excellently
summarized in~\cite{workshop,kadyk,vavra,cern1}, that suggests the 
wire chambers lifetime could be affected, and in some cases 
dramatically, by the nature and purity
of the gas mixture, materials used in contact with the gas,
by different additives, and trace contaminants.

While the specific reactions responsible for wire chamber aging 
are extremely complex, 
a qualitative description of the aging phenomena
in different gases could be obtained
from similarities between chemical processes in wire chamber
avalanches and those
that occur in the better-understood low-pressure 
rf-plasmas~\cite{kadyk,vavra,gcms,chemmod,hess}.
During wire avalanches many molecules break up 
by collisions with electrons, de-excitation of atoms, 
and UV-photon absorption processes. 
Whereas most ionization processes require electron energies 
greater than 10~eV,
the breaking of the chemical bonds in molecules and formation of free
radicals requires only 3-4~eV, and could lead to a large
abundance of free radicals over the ions in the  wire avalanches.
Similar to plasma polymerization~\cite{yasuda}, 
a free-radical polymerization 
seems to be the most appropriate mechanism of the wire chamber aging.
 Since free radicals are chemically very active they will either
recombine back to original molecules or other volatile species,
or form new cross-linked molecular structures, 
which could react further to produce chain-like polymers
of increasing molecular weight, thus lowering
the volatility of the resulting product.
 When the growing of polymerized chain becomes large enough
for condensation to occur, it will diffuse to the
electrode surface.
The polymer deposition mechanism can be 
conceived as a phenomenon that
occurs whenever the gaseous species fails to bounce back on 
collision with a electrode surface, including a surface of particles 
already formed in the gas discharges.
Initially the polymer could be held to the surface very weakly,
unless 
some additional chemical reaction takes place between the polymer 
atoms and atoms of the wire material. 
Furthermore, many free radicals are expected to have 
permanent or induced dipole moments; therefore,
electrostatic attraction to a wire could
play a significant role in the polymer deposition process.
Thus, aging effects in gaseous detectors would lead to 
deposition of thin, very high quality polymers on the anode wires, 
and would result in a loss of the gas gain and degradation
of ionization measurements.
Deposits on cathodes can induce discharges by secondary
electron emission (or Malter effect~\cite{malter}), and would lead to 
chamber breakdown.

Traditionally, the aging rate was parameterized as a normalized gas gain 
loss~\cite{kadyk}:

\[ R = - \frac{1}{G} \frac{dG}{dQ} ( \%~per~C/cm ) \]

where $G$ is an initial gas gain, 
$dG$ is the loss of gas gain after collected charge $Q$ per
unit length.
 The parameterization used above implicitly assumes that the gain drop
depends linearly on the charge collected per unit length of the wire, and 
that the aging rate is only a function of the total collected charge,
independent of gas gain and radiation intensity. In reality, this
assumption is not fulfilled.
 Several examples in the literature~\cite{juri,kott,opensh1,algeri,cern1} 
clearly indicate that the degree of aging for a given 
radiation dose could also depend upon the mode of operation,
being larger for smaller current densities, other conditions
being held constant. These observations raise a question on the
relevance of aging results, obtained in accelerated aging tests
with extremely high current densities, extrapolated to  
normal conditions.
  Moreover, aging studies of ATLAS muon drift tubes,
have shown a strong dependence of the aging
rate on the mode of operation (irradiation rate, high voltage)
and size of the irradiated area~\cite{dissert}.
Furthermore, it has been found that
the aging rate in the HERA-B honeycomb tracker chambers varies
substantially, when changing from $\gamma$ irradiation
to heavy ionizing particles~(see section 3.1). 
 This paper presents evidence  that the 
aging rate for irradiation with 5.9~keV $X$-rays and 
100~MeV $\alpha$'s may differ by more than two orders of 
magnitude~(see sections 4 and 5). 
Furthermore, it was recently reported in~\cite{lyon} and 
it will be shown in a subsequent paper~\cite{hadr}, that
the rate of polymer deposition could also be affected by the 
mode of operation, size of the irradiated area and 
water addition.


 These results exclude the possibility to extrapolate aging
rates obtained in the laboratory setups by irradiating
a small region of the wire to the large areas of
irradiation in hadronic environment. Therefore, for each 
particular case, the final decision for long-term 
stability of operation should be taken only after obtaining similar 
performance in conditions as close as possible to real ones.

\subsection{ Results from wire chambers operation. }

Over the last few decades, an impressive variety of
experimental data has been accumulated and it is not feasible
to go through all published results. In addition, new 
results make doubtful the quantitative comparison of
aging properties, realized at very different conditions.
Therefore, we will present below only a brief summary of the most recent
data, focusing on the binary and ternary mixtures, containing
$Ar$, $CH_4$, $CO_2$ and $CF_4$.

There are a lot of reports 
that clearly indicate premature aging in $Ar/CH_4$ filled
wire chambers exposed to intense 
radiation~\cite{juri,tsarnas,smith,kollef,kwong}.
 The polymerization of $CH_4$ is attributed to the
hydrogen deficiency of radicals and their ability to make bonds 
with hydrocarbon molecules~\cite{kadyk,yasuda}, leading 
to the buildup of deposits
on the anode wires, which usually contain carbon and lighter
elements. 

Attempts were made to replace the organic quenchers with more
stable ones, such as $CO_2$. 
However, the gradual decomposition of $CO_2$ in $Ar/CO_2$ mixtures
also takes place and stable operation 
was found to be possible up to charges of 
0.5-0.7~$\frac{C}{cm~wire}$~\cite{vavra,atlas}. 
Furthermore, carbon dioxide is also known for its
carbon buildup, which occurs specifically at the 
cathode~\cite{co2}, however, this carbon layer
did not affect the performance of the drift
tubes~\cite{atlas}.
Another effect, which is enhanced in $Ar/CO_2$ mixtures 
compared to gases containing hydrocarbon quenchers,
is the spurious counting rate~(noise), that presumably comes
from the spikes on the wires. This noise
can be cured by the
addition of water vapor to the drift gas~\cite{atlas}.

The aging studies in the ternary mixtures $Ar/CH_4/CO_2$
indicate the dependence of the aging rate on the 
particular set of operating conditions,
in some cases being smaller for larger $CH_4$ content 
in the mixture~\cite{kott,atlasnote,sadr}. 
The stable operation far beyond 0.6~$\frac{C}{cm~wire}$ 
was observed for $Ar/CH_4/N_2/CO_2$ (94:3:2:1) 
+ 1200~$ppm$ of $H_2 O$ mixture, however,
some conditions where found were only a tenth of 
nominal lifetime was achieved~\cite{kollef}. Moreover,
for the $Ar/CH_4/N_2/CO_2$ mixture, measurements have shown 
that their sensitivity to aging decreases with 
decreasing $CH_4$ and increasing $CO_2$ content.

The use of gases having $CF_4$ as a component was found to be  
attractive in the plasma processes, since the $CF_4$
is an ideal source for a variety of reactive neutral and ionic 
fragment atoms and molecules,
and especially neutral fluor, which is a  
desirable active species in etching processes~\cite{mogab}.
The $CF_4$ molecule is 
relatively inert in its electronic ground state, but
all excited electronic states of $CF_4$ and $CF_4^{+}$ ion
lead to dissociation.
Below the onset of electronic excitations, which is rather
high - 12.5~eV, collisions of electrons with $CF_4$ molecule 
lead to elastic scattering, vibrational excitation and 
dissociative attachment. Above this energy, 
dissociation of the $CF_4$ into neutrals or charged
fragments becomes significant~\cite{cf42,cf41}.
Actually, in plasma environment $CF_4$-based gases are used for 
both etching and deposition processes, 
the distinction being made by the gas and its amount with which $CF_4$ is mixed. 
The balance between these processes is called 'competitive 
ablation and polymerization'.
In general, the addition of
oxygenated species shifts the chemistry of $CF_4$ plasmas toward
etching, while the addition of hydrogenated species shifts the 
chemistry toward polymerization~\cite{chemmod,yasuda,kushner,winters}.
 For example, the polymer formation by $CF_4$ plasmas 
in the presence of $CH_4$ was observed~\cite{yasuda}.

 In the wire chamber operation, the usage of $CF_4$ is 
also quite controversial. 
 Many studies have demonstrated excellent aging properties, 
up to 10~$\frac{C}{cm~wire}$,
of $CF_4/i C_4 H_{10}$ (80:20) avalanches~\cite{kadyk1,opensh}, 
which also 
has an ability to etch both silicon-based~\cite{kadyk,opensh} and
hydrocarbon~\cite{opensh} deposits from the previously aged 
gold-plated wires. 
However, another proportions of $CF_4$ 
and isobutane could totally change the chemical reactions on the surface
of the anode wire.
Heavy carbonaceous deposits were formed on the gold-plated
wires irradiated in the mixtures $CF_4/i C_4 H_{10}$ (95:5) and
$CF_4/i C_4 H_{10}$ (20:80). The absence
of fluorocarbon deposits on the anode wires 
indicates that the deposits were formed only from $i C_4 H_{10}$,
without incorporation of $CF_x$ fragments.
Details of the chemical model and aging results in the 
$CF_4/i C_4 H_{10}$ are presented elsewhere~\cite{chemmod}.
Another gas observed to form heavy carbonaceous anode deposits
was $CF_4/C_2 H_4$(95:5)~\cite{chemmod}. At the same time,
the 4 $\%$ addition of $CF_4$ has been shown to inhibit anode damage
in $Ar/C_2H_6$(50:50) mixture up to the doses of
1.5~$\frac{C}{cm~wire}$~\cite{opensh1}. Aging studies have been
also performed with the $CF_4/CH_4$ (90:10) gas mixture 
and no deterioration was found up to 1.9~C/cm of the wire~\cite{d0}.
  Finally, it should be noted, that most of the aging studies in the
$CF_4$/hydrocarbon mixtures, were performed by irradiating a small region of the wire 
under well controlled laboratory conditions.
 At the same time, several painful experiences with aging 
are known from the high energy experiments.
The D0 muon drift chambers, 
filled with $Ar/CF_4/CH_4$(80:10:10) undergo a fast aging when
operated in a radiation environment. 
Vapors from a glue used in the construction were deposited on the wires
in a sheath, with the deposition rate proportional to accumulated charge~\cite{d0note}.
Furthermore, operation in the high-rate hadronic HERA-B environment
showed that the aging rate could depend on the type of 
ionization and size of the irradiated area~\cite{hohlmann,kolanoski}.
The severe anode and cathode aging effects
were observed in the honeycomb chambers tested in the
HERA-B conditions with $CF_4/CH_4$(80:20) and $Ar/CF_4/CH_4$(74:20:6) mixtures
after the radiation dose of several $mC$ per $cm$ of the wire.
This effect was surprising since the
chambers were proved to be immune a very large $X$-ray doses, up to
the 5~C/cm of the wire. 
 
 The aging properties of the gas mixtures 
$Ar(Xe)/CF_4/CO_2$, which by analogy with plasma experiments
should provide etching environments in the wire avalanches,  
have been also widely investigated over the past years.
The radiation tests of the ATLAS straw tubes using 
$Xe/CF_4/CO_2$ mixture showed no signs of aging 
for a integrated charge of up to 8~$C/cm$~\cite{akesson,straw}. 
However, the
detailed examination of anode wires and straw walls (cathode)
after irradiation had shown evidence for the etching
processes on both surfaces~\cite{romaniouk}.
 Another interesting phenomena was found in straw tubes
irradiated under exceedingly high current densities in $Xe/CF_4/CO_2$ mixture.
After an accumulated charge of 9~$C/cm$ the 
`anode swelling effect' was observed,
where destruction of gold coating of the gold-plated tungsten wires, 
followed by the tungsten oxidation~\cite{kriv1,kriv2}.
 Recently, it was also reported~\cite{hohlmann,kolanoski}, that 
under a high radiation intensity in $\afo$ (65:30:5) mixture, 
gold-plating of tungsten wires
can be etched away completely exposing the tungsten. 
However, the addition of 
water to the $\afo$ (65:30:5) was found to reduce the rate of 
etching processes, and the honeycomb chambers were able to survive 
radiation doses of up to 2~$\frac{C}{cm~wire}$.
The chemical etching processes in the $\afo$ 
are likely to be responsible for the
enhanced removal of the diamond coating observed in 
MSGC-GEM chambers~\cite{beauty}. 

Amazingly, extremely rapid aging has been observed 
for pure $CF_4$ and in $Ar/CF_4/O_2$(50:40:10) mixture,
which were expected to have strong etching abilities~\cite{chemmod,coyle}.
 In the former case, the aging process was related to chemical processes 
at the cathode, where trace fluorocarbon deposits were found.
 This type of aging phenomena was found to be cathode-material-dependent,
and resulted in a loss 
of gas gain rather than in a self-sustained discharge.
For the latter case, although $C$, $O$, and $F$ elements were observed 
on the anode wire, it is likely that the rapid aging was as a result of
the same (cathode phenomenon) observed in pure $CF_4$. 
The fluorination 
of the anode wires, found after exposure in $Ar/CF_4/O_2$(50:40:10) avalanches,
can cause anode aging in addition to cathode effects~\cite{chemmod}.
 A further phenomena due to the consistency of the cathode surfaces
was observed in ATLAS aluminum drift tubes operated in 
$Ar/CF_4/N_2/CO_2$ (94.5:0.5:2:3). Here, no change in gas gain 
for the tubes was found up to an accumulated charge of 
5~$\frac{C}{cm~wire}$.
However, the increased count rate at the beginning of the tests,
due to afterpulses in this gas at high gains
decreased with the time to the level, when no significant
afterpulsing was observed. 
This reduction was suggested to be due to the increased layer
of fluor and oxygen found at the cathode~\cite{dissert}.
 Similar effects were also found during the long-term irradiation of drift tubes 
with stainless steel cathodes in pure $CF_4$~\cite{denisov}.

These aging results clearly demonstrate that
the presence of large amounts of $CF_4$ in the mixture does not
necessarily ensure good aging properties. 
Moreover, perfluorocarbons represent the most extreme case of ablation
competing with polymer formation.
Therefore, all attempts
to improve the aging performance of the mixtures by adding $CF_4$
must be tested in the actual environment.

An abundance of literature exists describing the effect of certain additives
with oxygen-containing groups, like water, alcohol, methylal, oxygen.
These components, added to the mixture in small concentrations,
usually (but not always) extend the chamber lifetime,
especially against the polymerization of hydrocarbons.
 Of special interest is the addition of water vapors,
that has been found to suppress effectively polymerization in
wire chambers~\cite{kollef,charm}, to prevent Malter breakdowns~\cite{straw1},
or even to restore the original operation in aged 
counters~\cite{algeri,argus}; a few hundred to a few thousand $ppm$ of water
are generally used. The most natural mechanism by which water impedes Malter 
breakdowns and significantly improves the analog response of the
aged counters comes from the fact that water tends to increase the
conductivity of the damaged electrodes~\cite{kadyk,algeri}. 
In addition, water could
effectively suppress the secondary photon-mediated phenomena, if
the UV photons from carbon excitations are responsible for the
photoionization at the cathode~\cite{vavra2}. At the same time, the
exact mechanism by which water stops or reduces the aging rate 
in wire chambers is not clearly understood. One possible explanation
could be gained from the similarities with plasma chemistry,
where water acts like a blocking agent of the chain growth
mechanism, reacting with polymer precursors, thereby decreasing the
concentration of free radicals available for polymerization~\cite{yasuda}.
 Another explanation is that water, due to its large dipole moment,
could effectively slow-down electrons in the avalanches 
making the mixture `cooler'. In this case, aging effects resulting from free
radical polymerization would be reduced, since fewer
radicals would be formed in the wire avalanches. 
 Finally, it is well established, that
aging effects in the wire chambers, even when they are entirely due to
polymerization of gas mixture components can be accelerated by the
discharges~(glow discharges of sparks) 
and photoelectric feedback~\cite{vavra2,atac}. 
These effects are strongly dependent on the quality and type of the 
electrodes: surfaces of anode wire or cathode could be rather 
imperfect, thus, `triggering' polymerization in wire chambers. 
Known remedies for aging in this case are also water and alcohols,
which may prolong the lifetime considerably.
Furthermore, aging can be 
initiated by the trace contaminants such as $Si$, halogens~(other than $CF_4$),
sulfur, phthalats, benzine rings, which are either initially present in the 
gas, or result from outgassing of solid materials in contact with gas.
The detailed summary of the influence of
commonly used materials on aging properties 
may be found in ~\cite{kadyk,romaniouk}. 
When constructing large systems, 
one should before to certify the gas purity and chemical reactivity
of various materials in order to avoid the presence of 'bad' molecules 
in contact with the active gas volume.

\section{ Aging results from laboratory tests.}

 We have carried out a laboratory study of the aging properties
of $\afh$ (74:20:6) mixture, when exposed to an intensive $Fe^{55}$ source, 
and of $\afh$ (67:30:3) and $\afo$ (65:30:5) mixtures, both irradiated
with a $Ru^{106}$ source, up to 
a total radiation dose of 2~$C/cm$ of the wire.
 In each test, a single cell  
of the proportional tube chamber was irradiated through 
a hole in the cathode wall, closed  with thin $Al$ foil,
in the center of the 50~cm long cell.
 The irradiated area was limited by a collimator
to about 30~mm of the wire for the $Fe^{55}$ source 
and to 10~mm for the $Ru^{106}$ source.
While the $Fe^{55}$ 5.9~keV $X$-rays produce
an ionization cluster in the chamber cell, 
corresponding to $\sim$~200 primary electrons, the 
$Ru^{106}$ source provides 3.54~MeV electrons
from the decay of $Rh^{106}$, which leaves an ionization track
with $\sim$~90 primary electrons per cm. 
 The degree of aging  was determined quantitatively by
comparing $Fe^{55}$ spectra, measured at the irradiated and reference
positions of the wire, using the same electronics chain.
This method removed problems 
arising from the $Fe^{55}$ peak position movement due to changes in 
the pressure, the temperature, and gas mixture itself.
 The total collected charge was
determined by integrating continuously the recorded anode current.
The gas flow rate was 1~liter/hour for all tests.

\setlength{\unitlength}{1mm}
\begin{figure}[bth]
 \begin{picture}(160,110)
 \put(-15.0,-20.0){\includegraphics{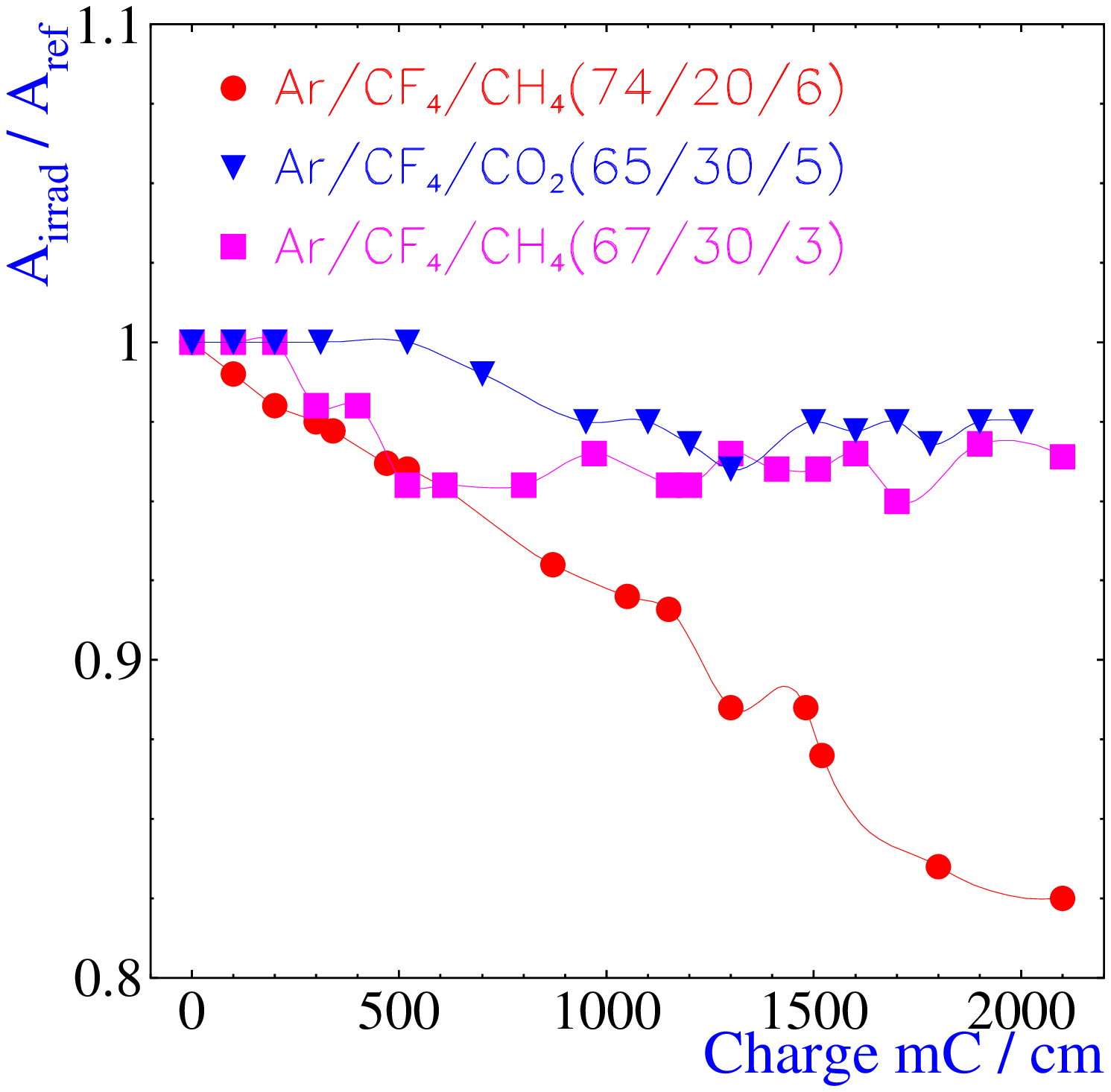} }
 \put(70.0,-20.0){\includegraphics{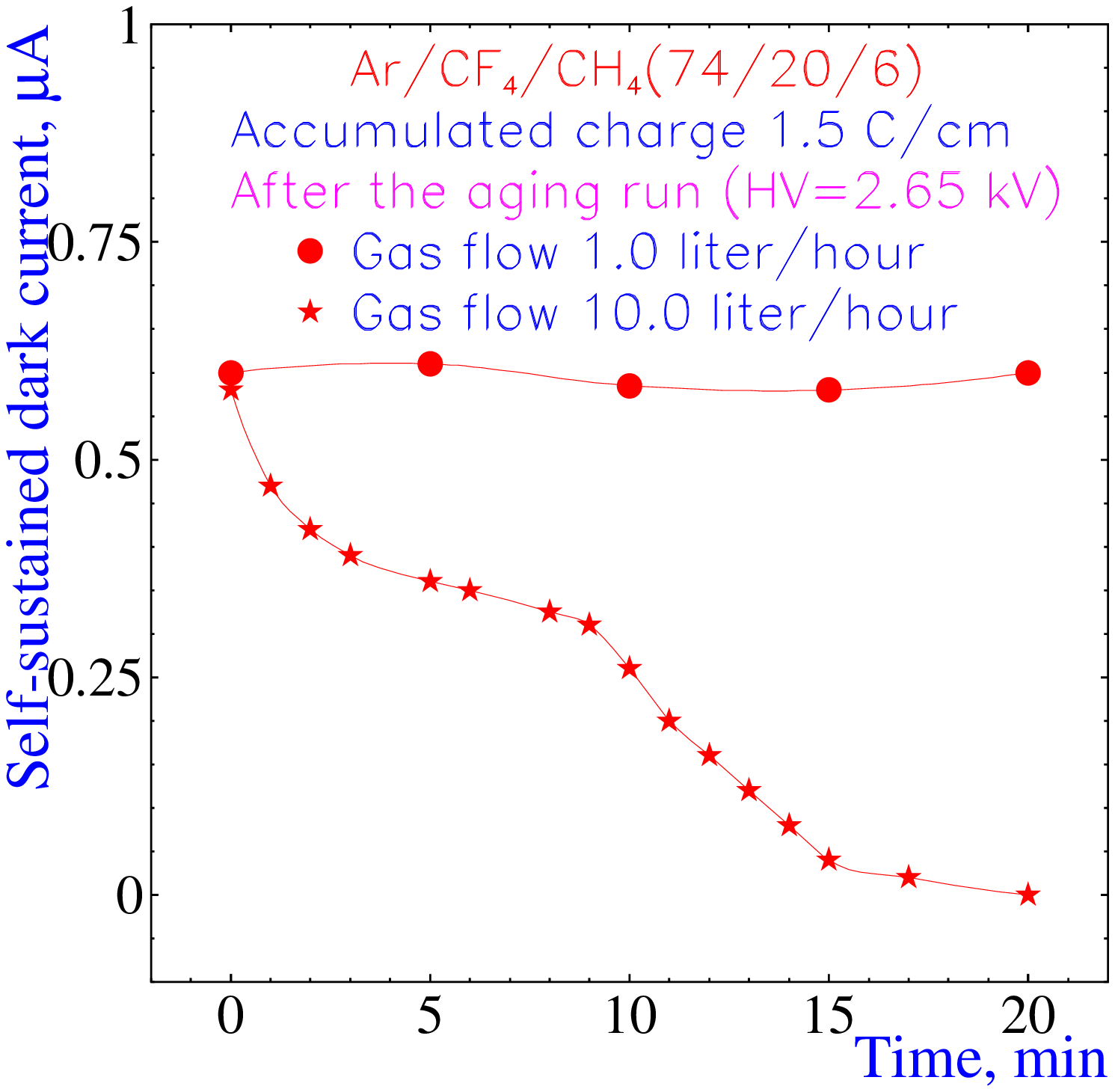} }
 \put(11,20){Figure 6: Ratio of $Fe^{55}$ peak positions }
 \put(12,14){ at irradiated and reference spots as a}
 \put(14,8){ function of the accumulated charge. }
 \put(102,20){Figure 7: Measured dependence of a}
 \put(106,14){self-sustained dark current as a }
 \put(106,8){ function of gas flow, after the }
 \put(104,2){ removal of the irradiation source. }
 \end{picture}
\end{figure}

A standardized set of conditions was established for the $Fe^{55}$
irradiation of the $\afh$ (74:20:6) mixture, corresponding to a current of
0.7~$\mu A/cm$ at a HV=2.65~kV. 
For irradiation with the $Ru^{106}$ source, the high voltages for
operation in $\afh$ (67:30:3)~($\afo$ (65:30:5)) mixtures
were set to 2690~V (2790~V), corresponding to current densities
of 1.2~$\mu A$ (0.9~$\mu A$).
 The different current densities for the 
irradiation of $\afh$ (67:30:3) and $\afo$ (65:30:5) mixtures
is attributed to the reduced source strength for the latter case,
due to the larger thickness of $Al$ foil used 
as a cathode above the irradiated region of the wire.
 The high voltages used during the aging studies already exceed
the proportional mode of gas amplification, 
therefore, events with higher pulse heights than those in the
proportional mode also appear at these voltages.


Fig. 6 shows the ratio of $Fe^{55}$ peak positions, measured 
at irradiated and reference spots as a function of the accumulated charge. 
 For the $\afh$ (74:20:6) mixture,
the relative mean amplitude $\frac{A_{irrad}}{A_{reference}}$
monotonously dropped to about 85~$\%$ of the initial value,
after the radiation dose of 2~$C/cm$ of the wire,
corresponding to $R \sim 8~$ $\%$ /~$C/cm$. 
Such a rate of gas gain losses could be acceptable  for the long-term
operation of the muon chambers in HERA-B.
However, after a collected charge $\sim$ 1~$C/cm$ 
significant self-sustained dark current 
appeared in addition to those normally associated with radiation.
 This dark current persisted even after the removal of
the radiation source, but it 
was totally quenched, either by a large increase in the gas flow rate, 
from 1~liter/hour to 10~liters/hour~(see Fig. 7), 
or by turning off the high voltage.
 This phenomena is associated with the polymer coating
and subsequent charge buildup on the cathode surface,
by a mechanism similar to the well known Malter discharge~\cite{malter}.

Irradiation with the $Ru^{106}$ source of 
$\afh$ (67:30:3) and $\afo$ (65:30:5) mixtures
led to $R$ values consistent with  zero~(see Fig. 6).
Such promising aging behavior make these gases suitable candidates
for use in the tube chambers. Unfortunately, as will be reported 
in section 5, the aging results, obtained in the high-rate
hadronic environment when irradiating a large area chamber differ 
significantly from those described above.

\section{ Aging studies in a 100~MeV $\alpha$-beam }

\subsection{ Aging in $\afh$(74:20:6) gas mixture.}

 After the tests with radioactive sources,
aging studies with the  $\afh$ (74:20:6) mixture were carried out 
by irradiating a single layer tube chamber 
in a 100~MeV $\alpha$-beam in Karlsruhe~(see Fig. 8).
 Due to the small range of the 100~MeV $\alpha$'s  
in aluminum, the thickness of the $Al$-wall 
was reduced to $\sim$ 200~$\mu m$.
 The $\alpha$-beam intensity was 
almost uniform over the beam area 8~$\times$~8~$cm^2$.
 This allowed a constant
irradiation of four 
wires fully, and, in addition, two
wires were exposed over half of their cell width. 
All wires were equipped for analog readout.  
Furthermore, for one of wires, which we will call a `reference wire',
two holes were made in the chamber wall - within the region
of irradiation and outside of it. Both holes
were closed with kapton foil.
For the `reference wire', it was possible
to measure the degree of aging by comparing the $Fe^{55}$
spectra at the two positions of the same wire.
 
A premixed gas mixture of $\afh$(74:20:6) 
was transported by stainless-steel tubes connected directly
to the tube chamber inlet and outlet. The gas flow was in serial
inside a chamber from one cell to another, 
with the flow rate about 6~liters/hour.
 The operating voltage was set to 2.35 kV, corresponding to a
gas gain $\sim 4 \times 10^{4}$, measured with $Fe^{55}$ $X$-rays.
 The radiation intensity varied during the aging studies 
by a factor of 3, thus excluding any possibility to
investigate aging behavior as a function of rate.
 Therefore, the current densities were in the range 
$\sim$~250~nA/cm up to 750~nA/cm.
By monitoring the chamber current, the total collected charge
was determined.

 Since the average energy losses of $\alpha$-particles 
in the chamber cell are much larger than those of $X$-rays, 
the same avalanche size as that for $X$-rays is obtained 
at a lower voltage. 
In practice, at an applied voltage of 2.35~kV,
the total charge released from $\alpha$-particles  
was measured to be a factor of 10-15 
larger than that corresponding to a 5.9~keV $X$-ray
localized energy deposition in proportional mode.
Moreover, the signals from $\alpha$'s at this voltage
were also accompanied by self-quenching streamer discharges, 
observed in several percent of the events at this voltage.
 This may be explained by the fact that since the transformation
from proportional to streamer mode depends upon the primary
ionization density~\cite{gang}, heavily ionizing particles should enter 
the self-quenching streamer mode at lower voltages than $Fe^{55}$ $X$-rays
and minimum ionizing particles.

\setlength{\unitlength}{1mm}
\begin{figure}[bth]
 \begin{picture}(160,110)
 \put(-40.0,-37.0){\includegraphics{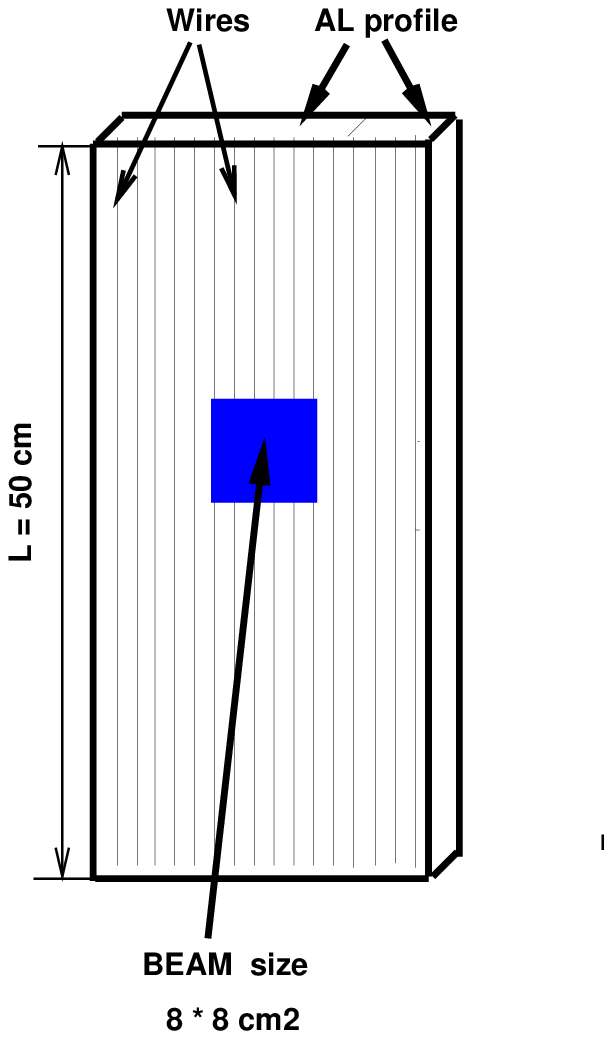} }
 \put(75.0,-20.0){\includegraphics{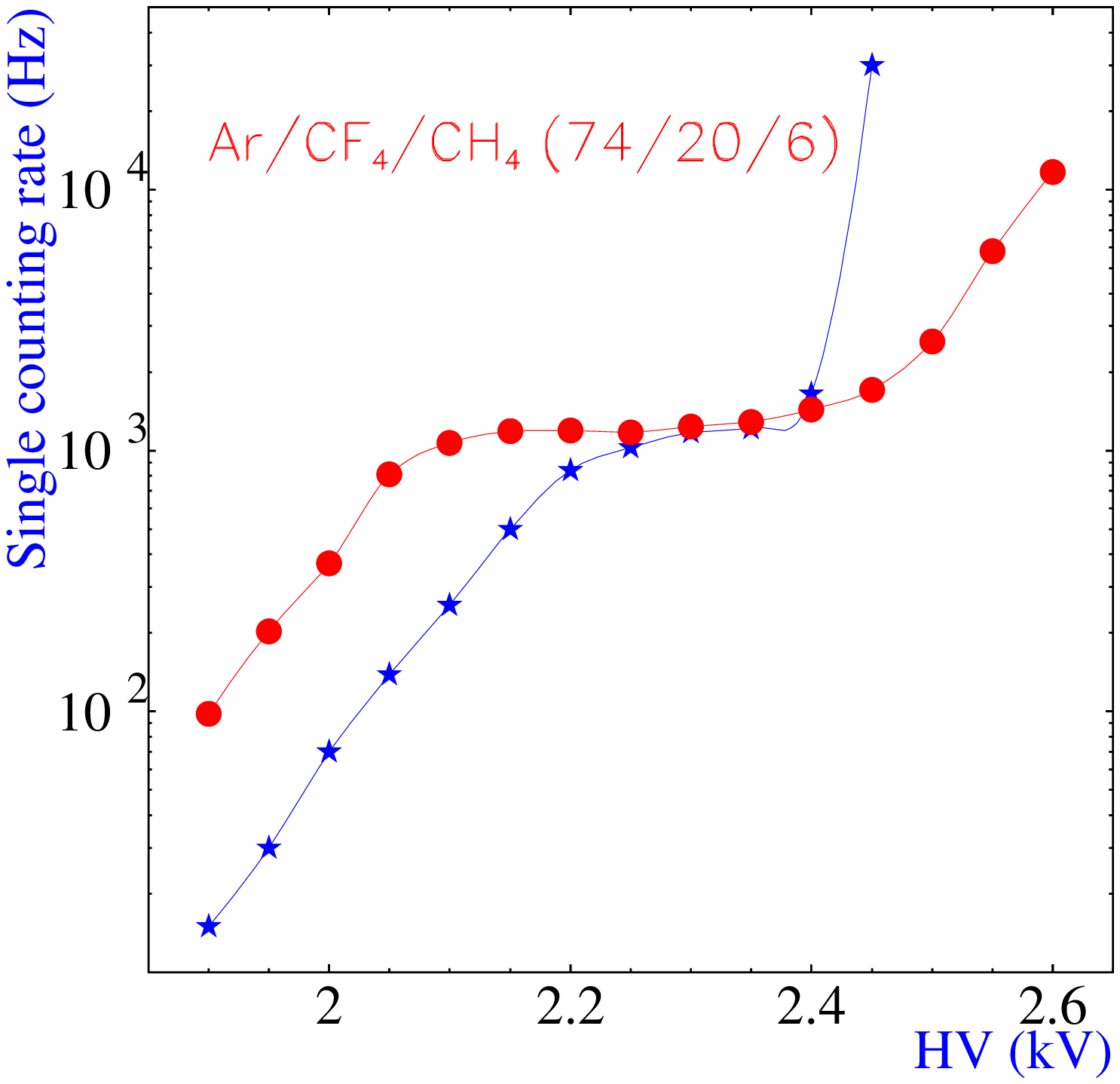} }
 \put(11,20){Figure 8: Sketch of the tube chamber }
 \put(12,14){ used in the $\alpha$-beam in Karlsruhe. }
 \put(102,20){Figure 9: Singles rate curves for the  }
 \put(102,14){ `reference wire' in the irradiated region  }
 \put(102,8){ (stars), after exposure to 60~$\frac{mC}{cm~wire}$,}
 \put(108,2){ and outside of it (circles). }
 \end{picture}
\end{figure}

After the first 24 hours of irradiation, which resulted in a 
collected charge of $\sim$~50~$\frac{mC}{cm~wire}$,
the status of the `reference wire' was checked 
by recording the $Fe^{55}$ pulse-height spectra.
Severe anode aging was observed for this wire
in the irradiated area, corresponding to a normalized gas gain
loss value $R \sim 750~\%~per~C/cm $, while the $Fe^{55}$
spectra in the non-irradiated region remained unchanged.
The downward shift of the peak value was accompanied by a
distortion and broadening of the $Fe^{55}$ pulse-height spectrum.
 After the next 6 hours of exposure, when an additional 
10~$\frac{mC}{cm~wire}$ were collected,  
a steady decrease of the  
gas gain for the `reference wire' has been found, and the $R$ value 
became even slightly higher $\sim 850~\%~per~C/cm $.
 Furthermore, during these hours,
sizeable increase in the single counting rate,
above the actual $\alpha$-beam intensity,
was observed from several of the irradiated wires. 
Since that time, the irradiation of the `reference wire' was 
halted, but, the aging studies with all other wires were continued 
until the collected charge 
had reached a value of 280~$\frac{mC}{cm~wire}$.
Two wires, irradiated over half of their cell width, accumulated
half of this charge - 140~$\frac{mC}{cm~wire}$.
The aging behavior of the wires during irradiation was quite similar:
large increase in the single counting rate, as a result of
spark discharges,
was observed from all wires at an applied voltage of
2.35~kV, together with the monotonic
drop in the chamber current.


\setlength{\unitlength}{1mm}
\begin{figure}[ph]
 \begin{picture}(160,200)
 \put(40.0,160.0){\includegraphics{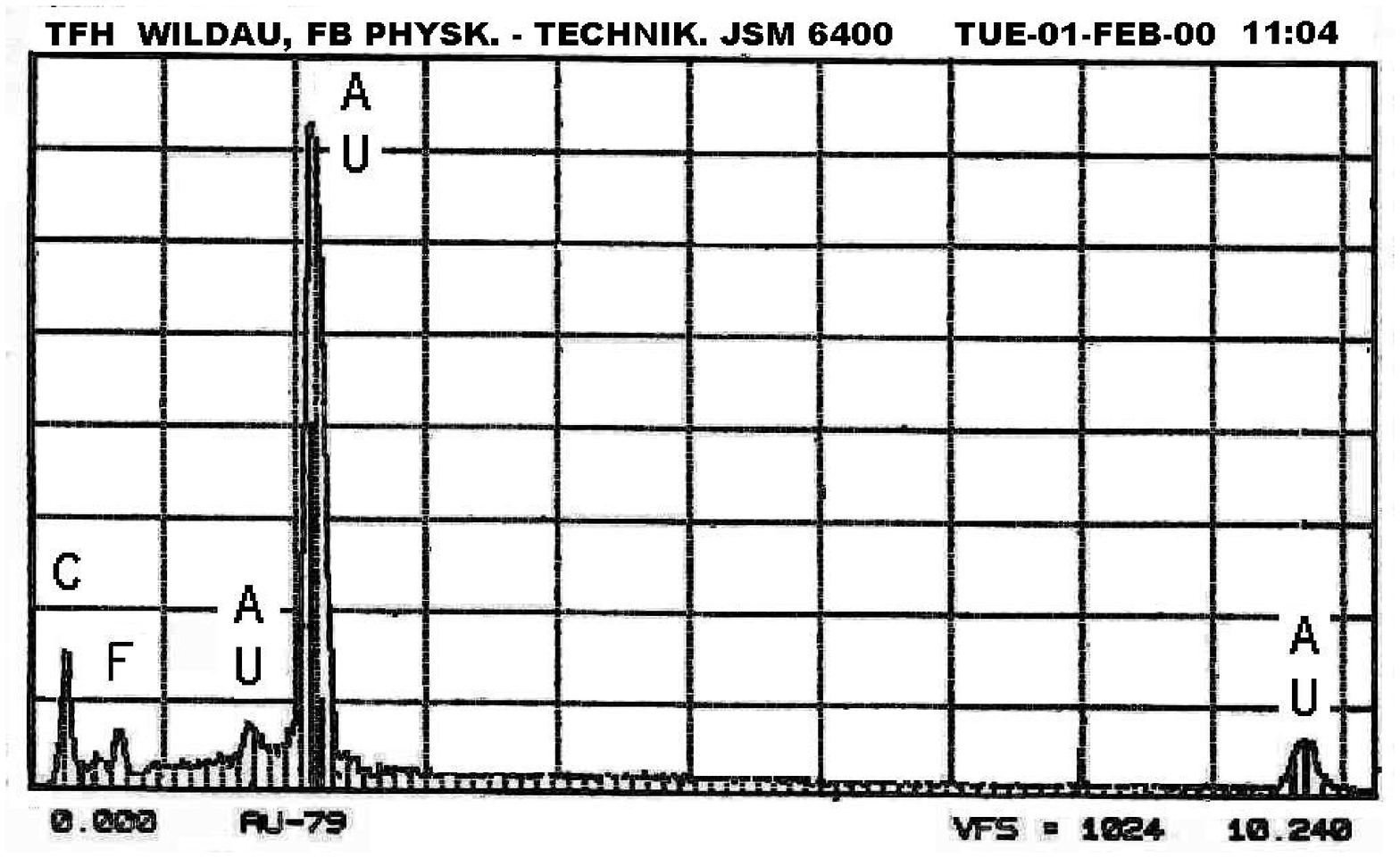} }
 \put(40.0,85.0){\includegraphics{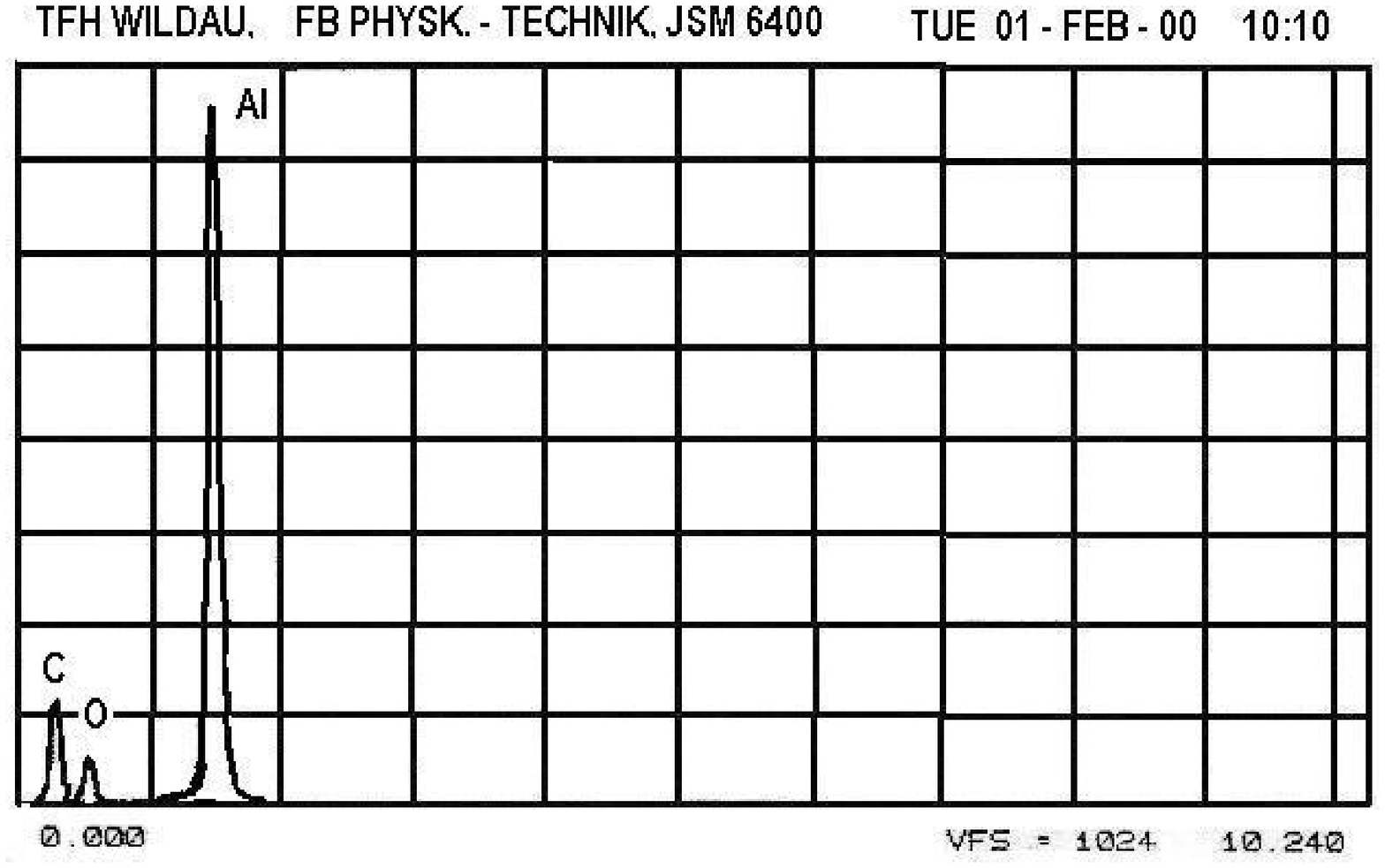} }
 \put(10.0,15.0){\includegraphics{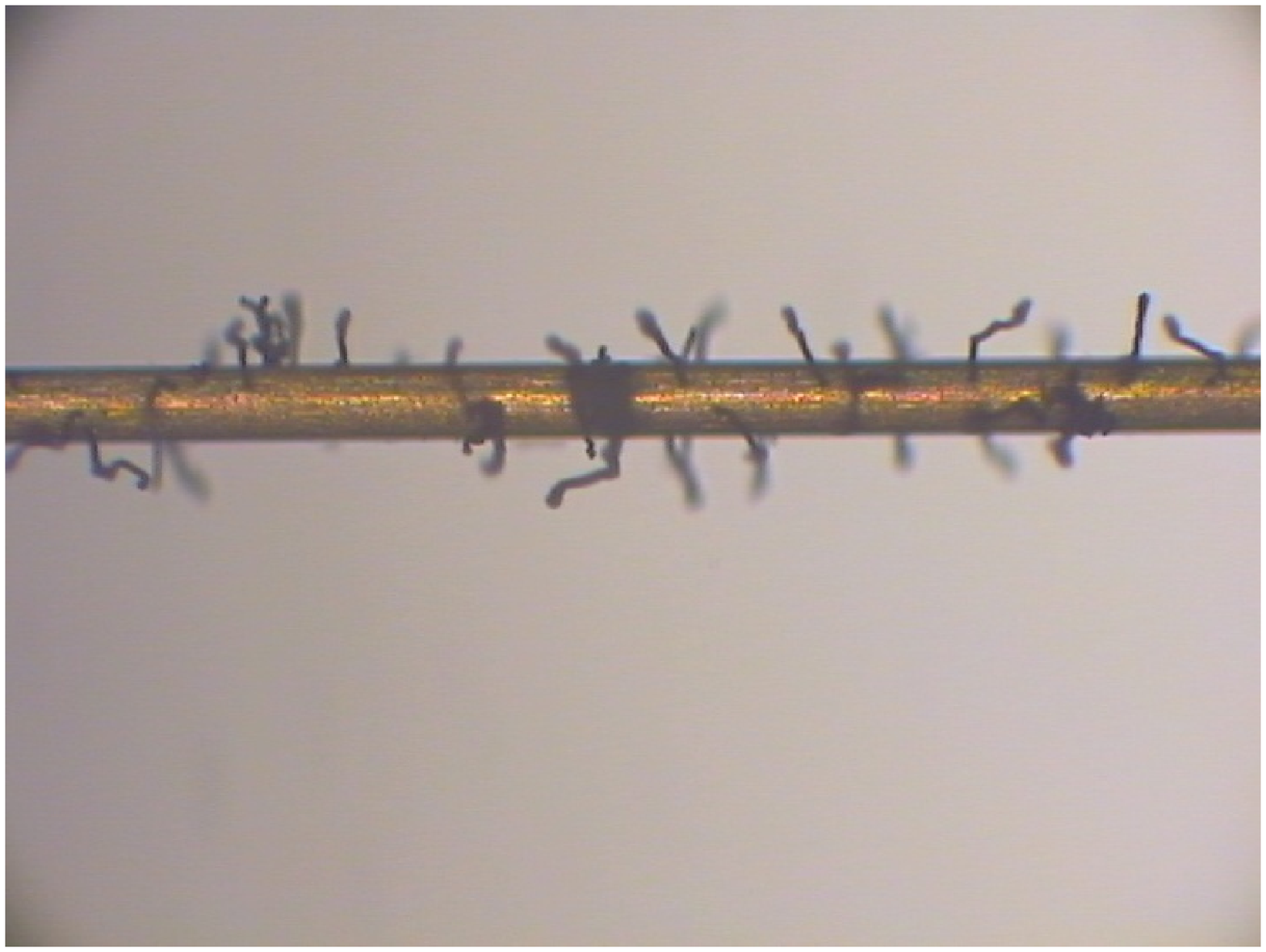} }
 \put(93.0,3.0){\includegraphics{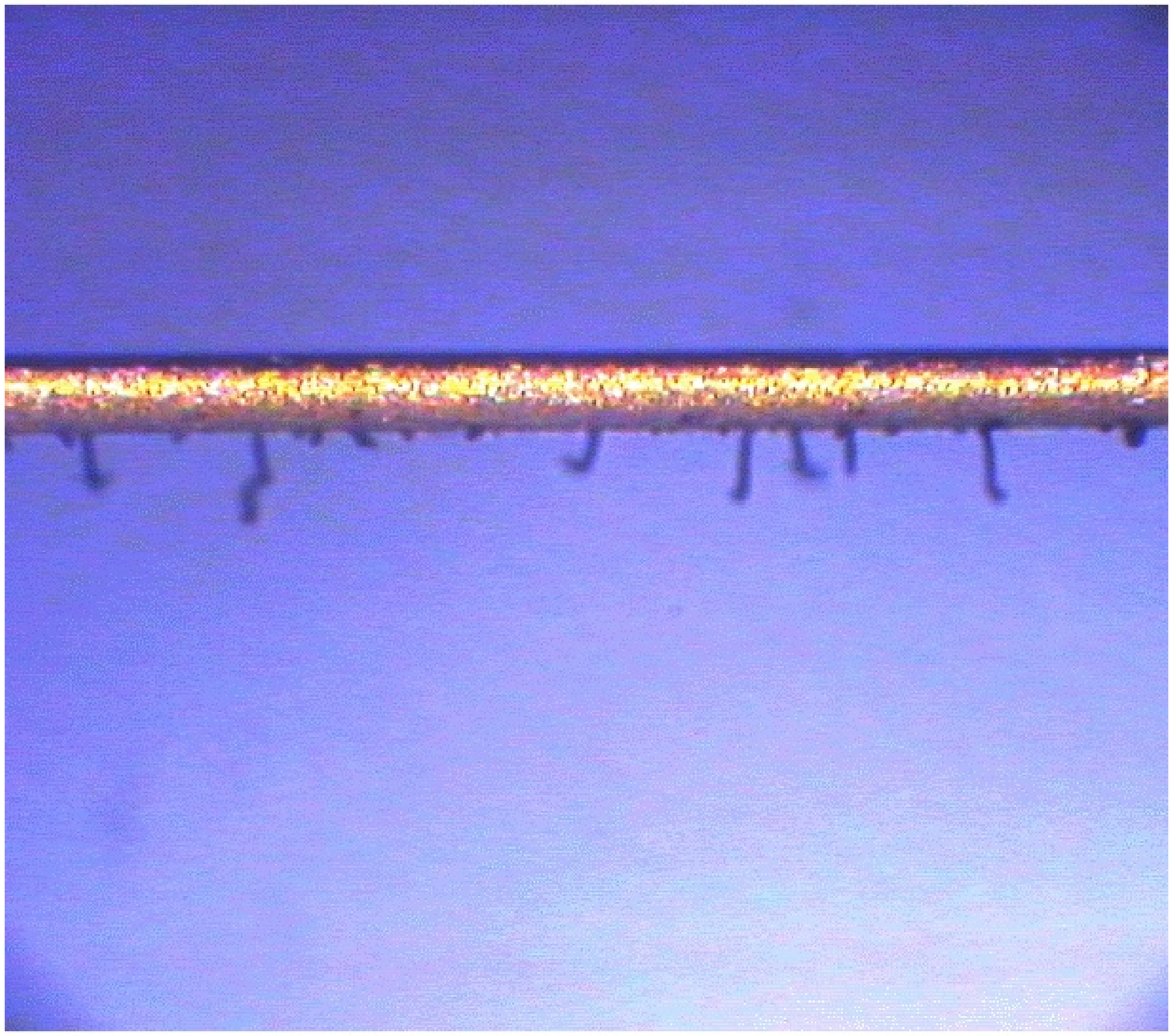} }
 \put(15,153){Figure 10: SEM of the 'reference wire' after the accumulated charge of 60~$\frac{mC}{cm~wire}$.}
 \put(15,147){ $C$ and $F$  
elements were identified together with Au signal from the wire material.}
 \put(15,83){Figure 11: SEM of the cathode surface composition in the irradiated region }
 \put(35,77){ of the wire after exposure to 280~$\frac{mC}{cm~wire}$. }
 \put(13,8){Figure 12: Typical deposits on the   }
 \put(9,2){ anode wire after exposure to 280~$\frac{mC}{cm~wire}$.}
 \put(98,8){Figure 13: Wire irradiated over half }
 \put(98,2){ of its cell width. Deposits built}
 \put(98,-4){up on one side of the wire towards }
 \put(98,-10){all of the primary electrons drifted. }
 \end{picture}
\end{figure}

After the aging studies in Karlsruhe, the chamber characteristics
have been studied on the 3~GeV electron beam with intensity
$\sim$~1~$kHz/cm^{2}$.
Fig. 9 shows the singles rate curves for the `reference wire'
in the irradiated region, after exposure to 
60~$\frac{mC}{cm~wire}$, and outside of it.
 From these curves it is seen that, as result of irradiation, 
there is a shift upwards of the 
lower end of the efficiency plateau~(showing a decrease of gain
at a given voltage) with a
consequent reduction in effective plateau length. 
 All other irradiated wires were also studied on the electron beam.
Wires that have been exposed to a charge
of 280~$\frac{mC}{cm~wire}$ were found to be heavily damaged,  
showing an efficiency at the level of ~$\sim$1~$\%$-10~$\%$
for the nominal operating high voltage 2.35~kV.
 In addition, for all damaged wires the breakdown knee
in the singles rate curves significantly displaced towards lower voltages.
This effect can also be seen for the `reference wire' in Fig. 9. 
For the damaged region of this wire,
the breakdown point occurs
at HV=2.45~kV, where the single counting rate increases abruptly.
At the same time, whenever the electrons illuminates regions of the 
wire that have not been previously irradiated, no actual difference
in the singles rate curves (before and after irradiation) shows up.

After the beam tests, 
the chamber was opened for inspection and both
anode wires and cathodes surfaces were analyzed, using 
microscope and later scanning electron microscopy~(SEM).
 Observation of all damaged wires   
under a microscope show vertically structured black deposits  
or 'whiskers'~(see Fig. 12), distributed
randomly within the irradiated area only,
and of up to lengths of 50~$\mu m$ from the wire surface. 
The 'whiskers' were found to be very fragile,
and easily broken if touched.
 Of special interest, there were two wires at the boundaries of the
irradiated area where only half of the cell width was exposed 
to $\alpha$'s. These wires showed significant asymmetries in 
the 'whiskers'~(see Fig. 13) with 
deposits built up on one side of the anode wire towards
all of the primary electrons drifted. 
 The observed asymmetries in the anode deposits indicate that 
even for high ionization densities produced by $\alpha$-particles,
the operating parameters (high voltage, gas mixture) 
led rather to avalanches  
well localized on one side of the anode wire rather than to avalanches 
whose spread was large enough to surround the wire completely. 

 Examination of all damaged wires by SEM
revealed deposits, containing carbon and fluorine 
as the only detectable elements,
thus confirming the presence of a polymer coating. 
Moreover, carbon and fluorine signals were also identified in the
irradiated regions free of 'whiskers',
together with Au signal from the wire material.
 Fig. 10 shows, as an example, the SEM of the `reference wire' after the 
accumulated charge of 60~$\frac{mC}{cm~wire}$.
 New wires and wires which were exposed to the irradiated gas,
but not irradiated themself, showed peaks only from Au.
Unfortunately, this method is unable to detect hydrogen,
so hydrocarbons registered as a carbon alone and are also 
trapped in the polymer coating.
 SEM analysis of the cathode surface composition  
in the irradiated region of the wire after exposure to
280~$\frac{mC}{cm~wire}$ was also performed.
Translation of peak areas 
into atomic abundances indicates that the top layer of the 
cathode consists mainly
of $Al$($>80~\%$) with some traces of carbon and oxygen~(see Fig. 11).

 A direct means of investigating the radiation-initiated spark discharges
was to separate anode and cathode effects by combining new and aged
portions of the tube chamber. When the cathodes of the tubes
that had been previously aged were restrung with the new anode wires,
the breakdown knee in the singles rate displaced to the higher voltages.
Therefore,
one reason for the discharges are some local
deposits on the anode wires~(conductive `whiskers').
It was reported by Holland~\cite{book3}, that the
properties of polymer films can be electrically
conductive or insulating according  to the carbon to hydrogen ratio
and a degree of cross-linking.
 If a coating is primarily carbon, 
rather than normal polymers, the conductivity will be much higher.
At the same time, the energy and 
amount of particle bombardment on the surface can very easily change
the composition and therefore conductivity. 
Hydrogen can be easily
removed by ion bombardment, so heavy bombardment leads to a more
carbonaceous than polymer-like deposit, which may be relevant
for the case of irradiation with $\alpha$-particles.


%




 


\section{ Summary and outlook. }

 Experimental conditions at HERA-B impose very strong
requirements for the gaseous detectors.
 The high radiation load of the HERA-B experiment results in an accumulated
charge of up to 200~$\frac{mC}{cm~wire}$ in the tube chambers. The 
chamber signals, which are used to form the trigger, must be collected 
in less than the 96~ns between bunch crossings.

 A short description of the criteria which are relevant for the choice of 
gases in the HERA-B muon detector are discussed and 
the tube chamber characteristics 
studied for several $Ar/CF_4$-based mixtures are reported.
However, the main criteria for the choice of the gases are the aging properties.
 In this paper, the brief summary of the most recent data
focusing on the binary and ternary mixtures, containing 
$Ar$, $CH_4$, $CO_2$ and $CF_4$ is presented.
Although, $CF_4$ gas has been widely used over the last ten years,
the results from it's usage in the drift tubes are quite controversial.
 The aging results reported in this paper 
demonstrate that the aging rate for irradiation with 5.9~keV $X$-rays and 
100~MeV $\alpha$'s may differ by more than two orders of magnitude.
 Despite the negligible gas gain loss  
measured after the long-term irradiation with $Fe^{55}$
$X$-rays, rapid aging effects observed with $\alpha$-particles 
completely ruled out $\afh$ (74:20:6) mixture as a  
candidate for operation in the HERA-B high radiation environment, 
where at least part of the ionization
is deposited by heavily ionizing particles.
  Moreover, 
from these results it is evident that from the accumulated charge alone, 
it is not possible to combine the data from the different radiation sources 
into one consistent model.
 Therefore, as long as the data from different aging tests can not be combined
properly, it is very difficult to give an extrapolation about the 
lifetime in the real detector.
 In order to find a link between 
results in different setups, we also performed
aging studies with $\afh$ (67:30:3) 
and $\afo$ (65:30:5) mixtures in the HERA-B environment,
under conditions as similar as possible to real ones.
 In a subsequent paper~\cite{hadr} we will report on the results from these tests, 
which indicate that the aging rate could also be affected by
the mode of operation, size of the irradiated area and water addition.


\section*{Acknowledgments.}

 We express our special appreciation for discussions with Dr. J. Va'vra.
 We would like to thank Dr. G. Bohm and Wildau
Politechnik Institute for the possibility to analyse the wires.
 We thank to K. Reeves and S. Aplin for reading and correcting this 
manuscript.

 This work was supported by the 
Deutsches Elektronen-Synchrotron (DESY), 
by the Alexander von Humboldt-Stiftung and Max Planck Research Award.



\begin{thebibliography}{99}

\bibitem{Proposal} T. Lohse $et~al.$, HERA-B collaboration,
An Experiment to Study CP Violation in the B System Using an 
Internal Target at the HERA Proton Ring, Proposal,
{\bf DESY-PRC 94/04} (1994). 
 
\bibitem{TDR} E. Hartouni $et~al.$, HERA-B collaboration,
An Experiment to Study CP Violation in the B System Using an 
Internal Target at the HERA Proton Ring, Design Report,
{\bf DESY-PRC 95/01} (1995). 

\bibitem{electronics} M. Buchler $et~al.$,
IEEE Trans. Nucl.Sci., {\bf NS-46 } (1999) 126-132.

\bibitem{scint} T.J. Sumner {\it et al},
IEEE Trans. Nucl. Sci., {\bf NS-29 (5)} (1982) 1410-1414.

\bibitem{start} S.A. Korff and R.D.Present, 
Phys. Rev. {\bf Vol.65 (9)} (1944) 274-282.

\bibitem{co2} V. Bawdekar,
IEEE Trans. Nucl. Sci {\bf NS-22 } (1975) 282-285.

\bibitem{N2} L. Colli, U. Fracchini,
Rev. Sci. Instr. {\bf Vol.23 (1)} (1952) 39-42.

\bibitem{book1} L.G. Christophorou, Atomic and Molecular 
Radiation Physics, (Wiley, New York, 1971)

\bibitem{fast1} L.G. Christophorou {\it et al}, 
Nucl. Instr. and Meth. {\bf A 163} (1979) 141-149.

\bibitem{fast2} L.G. Christophorou {\it et al}, 
Nucl. Instr. and Meth. {\bf A 171} (1980) 491-495.

\bibitem{armit} J. C. Armitage {\it et al},
Nucl. Instr. and Meth. {\bf A 271} (1988) 588-496.

\bibitem{cern} A. Peisert, F. Sauli,
CERN 84-08 (1984).

\bibitem{biagi} S. Biagi,
Nucl. Instr. and Meth. {\bf A 421} (1999) 234-240.

\bibitem{cf42} M.C. Borgade {\it et al}, 
J. Appl. Phys, {\bf Vol.80, (3)} (1996) 1325-1336.

\bibitem{mix1} T.Yamashita {\it et al}, 
Nucl. Instr. and Meth. {\bf A 317} (1992) 213-220.

\bibitem{transport} S.R.Hunter {\it et al}, 
J. Appl. Phys, {\bf Vol.58, (8)} (1985) 3001-3015.

\bibitem{christ} L.G. Christophorou {\it et al}, 
Nucl. Instr. and Meth. {\bf A 309} (1991) 160-168.

\bibitem{book2} L.G. Christophorou (ed.), Electron-Molecule
Interactions and Their Applications, vols. 1 and 2, 
(Academic Press, New York, 1984)

\bibitem{attach} W.S. Anderson {\it et al}, 
Nucl. Instr. and Meth. {\bf A 323} (1992) 273-279.

\bibitem{cf41} L.G. Christophorou {\it et al}, 
J. Phys. Chem. Ref. Data, {\bf Vol.25, No.5} (1996) 1341-1388.

\bibitem{quench1} L.G. Piper {\it et al}, 
J. Chem. Phys, {\bf Vol.59, (6)} (1973) 3323-3340.

\bibitem{quench2} J.E. Velazco {\it et al}, 
J. Chem. Phys, {\bf Vol.69, (10)} (1978) 4357-4373.

\bibitem{becker} U. Becker {\it et al}, 
Nucl. Instr. and Meth. {\bf A 315} (1992) 14-20.

\bibitem{grimm} O. Grimm, Diplomarbeit, 
University Hamburg, July (1998) (in german)

\bibitem{beck} M. Beck, Dissertation, 
University Rostock, June (1999) (in german)

\bibitem{mix2} T.Yamashita {\it et al}, 
Nucl. Instr. and Meth. {\bf A 283} (1989) 709-715.

\bibitem{gain1} S. Beingessner, R. Carnegie,
Nucl. Instr. and Meth. {\bf A 260} (1987) 210-220.

\bibitem{charpak} G. Charpak {\it et al},
Nucl. Instr. and Meth. {\bf A 99} (1972) 279-284.

\bibitem{workshop} Proc. Workshop on Radiation Damage to Wire Chambers, 
Lawrence Berkeley Laboratory (Jan. 1986) LBL-21170.

\bibitem{kadyk} J. A. Kadyk,
Nucl. Instr. and Meth. {\bf A 300} (1991) 436-479.

\bibitem{vavra} J. Va'vra, ref.~\cite{workshop}, pp. 263-294.

\bibitem{cern1} R. Bouclier {\it et al},
Aging of microstrip gas chambers: problems and solutions,
CERN-PPE 96-33 (1996).

\bibitem{gcms} J. Wise {\it et al},
IEEE Trans. Nucl. Sci {\bf NS-37 (2) } (1990) 470-477.

\bibitem{chemmod} J. Wise {\it et al}, 
J. Appl. Phys, {\bf Vol.74, (9)} (1993) 5327-5340.

\bibitem{hess} D. W. Hess, ref.~\cite{workshop}, pp. 15-24.

\bibitem{yasuda} H. Yasuda, Plasma Polymerization,
(Academic Press, 1985)

\bibitem{malter} L. Malter
Phys. Review, {\bf Vol. 50} (1936) 48-58.

\bibitem{juri} I. Juricic, J. Kadyk, ref.~\cite{workshop}, pp. 141-159.

\bibitem{kott} R. Kotthaus, ref.~\cite{workshop}, pp. 161-193.

\bibitem{opensh1} R. Openshaw {\it et al}, 
IEEE Trans. Nucl. Sci {\bf NS-36 (1) } (1989) 567-571.

\bibitem{algeri} A. Algeri {\it et al},
Nucl. Instr. and Meth. {\bf A 338} (1994) 348-367.

\bibitem{dissert} V. Pashhoff, 
Dissertation, University Freiburg, October (1999)

\bibitem{lyon} M. Titov, 
A Gaseous Muon Detector at the HERA-B experiment,
talk given at Nuclear Science Symposium and Medical Imaging Conference,
15-20 October 2000, Lyon, France 

\bibitem{hadr} M. Danilov, L. Laptin, I. Tichomirov, M.Titov, Yu. Zaitsev,  
Aging of gaseous detectors under high-rate irradiation with 
hadronic particles, in preparation.

\bibitem{tsarnas} N. Spielberg, D. Tsarnas
Rev. Sci. Instr. {\bf Vol.46 (8)} (1975) 1086-1091.

\bibitem{smith} A. Smith, M. Turner
Nucl. Instr. and Meth. {\bf A 192} (1982) 475-481.

\bibitem{kollef} M. Kollefrath {\it et al},
Nucl. Instr. and Meth. {\bf A 419} (1998) 351-356.

\bibitem{kwong} K. Kwong,
Nucl. Instr. and Meth. {\bf A 238} (1985) 265-272.
 
\bibitem{atlas} V. Paschhoff, M.Spegel, Aging studies for the
ATLAS MTD's using $Ar/CO_2$(90:10), ATLAS muon note-019 (1999).

\bibitem{atlasnote} I. Boyko {\it et al}, Aging of aluminum
drift tubes filled with $Ar/CO_2/CH_4$ (92:5:3), 
ATLAS muon note-089 (1995).

\bibitem{sadr} H. Sadrozinski, ref.~\cite{workshop}, pp. 121-129.

\bibitem{mogab} C. Mogab {\it et al}, 
J. Appl. Phys, {\bf Vol.49, (7)} (1978) 3796-3803.

\bibitem{kushner} M.J. Kushner,
J. Appl. Phys, {\bf Vol.53, (4)} (1982) 2923-2938.

\bibitem{winters} H.F. Winters {\it et al}, 
J. Appl. Phys, {\bf Vol.48, (12)} (1977) 4973-4983.

\bibitem{kadyk1} J. Kadyk {\it et al},
IEEE Trans. Nucl. Sci {\bf NS-37 (2) } (1990) 478-486.

\bibitem{opensh} R. Openshaw {\it et al},
Nucl. Instr. and Meth. {\bf A 307} (1991) 298-308.

\bibitem{d0} G. Alexeev {\it et al},
Technical design Report for the D0 Forward Muon Tracking
Detector Based on Mini-Drift Tubes, D0 Note 3366, (1997)

\bibitem{d0note} B. Baldin {\it et al},
Technical design of the central muon system, D0 Note 3365, (1997)


\bibitem{hohlmann} M. Hohlmann, The Outer Tracker of HERA-B, 
Proceedings of 8th Pisa Meeting on advanced detector,
Isola d'Elba'00, Italy (in press)

\bibitem{kolanoski}  H. Kolanoski, Investigation of Aging
in the HERA-B Outer Tracker Drift Tubes,
talk given at Nuclear Science Symposium and Medical Imaging Conference,
15-20 October 2000, Lyon, France 

\bibitem{akesson} T. Akesson {\it et al},
Nucl. Instr. and Meth. {\bf A 361} (1995) 440-456.

\bibitem{straw} ATLAS Technical Design Report, CERN (1997)

\bibitem{romaniouk} A. Romaniouk,
 Choice of materials for the constructio of TRT,
ATLAS Internal Note, INDET-98-211 (1998)

\bibitem{kriv1} G. Gavrilov {\it et al},
Aging investigation of ATLAS TRT straws, PNPI-preprint-2328 (1999)

\bibitem{kriv2} T. Ferguson {\it et al},
Possible new mechanism for anode wire aging
in gas filled detectors, PNPI-preprint-2331 (1999)

\bibitem{beauty} T. Zeuner, The MSGC-GEM Inner Tracker for HERA-B
Nucl. Instr. and Meth. A (Beauty'99 proceedings, in press).

\bibitem{coyle} J. Va'vra  {\it et al},
Nucl. Instr. and Meth. {\bf A 324} (1993) 113-126.

\bibitem{denisov} D.S. Denisov, 
On using $CF_4$ as a working gas for drift tubes (in russian),
IHEP-preprint-90-16 (1990)

\bibitem{charm} J. DeWulf  {\it et al},
Nucl. Instr. and Meth. {\bf A 252} (1986) 443-449.

\bibitem{straw1} J.Kadyk {\it et al},
Nucl. Instr. and Meth. {\bf A 300} (1991) 511-517.

\bibitem{argus} M. Danilov {\it et al},
DESY Internal Note 88-090, DESY (1988)

\bibitem{vavra2} J. Va'vra, 
Aging of gaseous detectors, SLAC-PUB-5207 (1990)

\bibitem{atac} M. Atac, ref.~\cite{workshop}, pp. 55-66.

\bibitem{gang} AN Ji-Gang {\it et al},
Nucl. Instr. and Meth. {\bf A 267} (1988) 396-407.

\bibitem{book3} L. Holland, Thin Film Microelectronics 
(Chapman and Hall Ltd, 1965) p. 157.


\end{thebibliography}
\end{document}